\documentclass[twocolumn]{aastex62}

\usepackage{graphicx} 

\graphicspath{{./}{figures/}}

\received{20 Nov, 2024}
\revised{8 Jan, 2025}
\accepted{9 Jan, 2025}
\submitjournal{AJ}
\shorttitle{ExoEcho. I. TTV From Hubble Observation}

\shortauthors{Ma et al}

\begin{document}

\title{ Exoplanet Ephemerides Change Observations (ExoEcho). I. Transit Timing Analysis of Thirty-Seven Exoplanets using HST/WFC3 Data}

%\title{ ExoEcho. I. Transit Timing Analysis of Thirty-Seven Exoplanets using HST/WFC3 Data}

%\title{Transit Timing Analysis of Thirty-Seven Exoplanets using HST/WFC3 Data}

\correspondingauthor{Bo Ma, Zhou Li}
\email{mabo8@mail.sysu.edu.cn; lizhou@bao.ac.cn}

\author{Xinyue Ma}
\affil{School of Physics and Astronomy, Sun Yat-sen University, Zhuhai 519082, China; {\it mabo8@mail.sysu.edu.cn}}
\affil{CSST Science Center for the Guangdong-Hong Kong-Macau Great Bay Area, Sun Yat-sen University, Zhuhai 519082, China}

\author{Wenqin Wang}
\affil{School of Physics and Astronomy, Sun Yat-sen University, Zhuhai 519082, China; {\it mabo8@mail.sysu.edu.cn}}
\affil{CSST Science Center for the Guangdong-Hong Kong-Macau Great Bay Area, Sun Yat-sen University, Zhuhai 519082, China}

\author{Zixin Zhang}
\affil{School of Physics and Astronomy, Sun Yat-sen University, Zhuhai 519082, China; {\it mabo8@mail.sysu.edu.cn}}
\affil{CSST Science Center for the Guangdong-Hong Kong-Macau Great Bay Area, Sun Yat-sen University, Zhuhai 519082, China}

\author{Cong Yu}
\affil{School of Physics and Astronomy, Sun Yat-sen University, Zhuhai 519082, China; {\it mabo8@mail.sysu.edu.cn}}
\affil{CSST Science Center for the Guangdong-Hong Kong-Macau Great Bay Area, Sun Yat-sen University, Zhuhai 519082, China}

\author{Dichang Chen}
\affil{School of Astronomy and Space Science, Nanjing University, Nanjing 210023, China}
\affil{Key Laboratory of Modern Astronomy and Astrophysics, Ministry of Education, Nanjing 210023, China}

\author{Jiwei Xie}
\affil{School of Astronomy and Space Science, Nanjing University, Nanjing 210023, China}
\affil{Key Laboratory of Modern Astronomy and Astrophysics, Ministry of Education, Nanjing 210023, China}

\author{Shangfei Liu}
\affil{School of Physics and Astronomy, Sun Yat-sen University, Zhuhai 519082, China; {\it mabo8@mail.sysu.edu.cn}}
\affil{CSST Science Center for the Guangdong-Hong Kong-Macau Great Bay Area, Sun Yat-sen University, Zhuhai 519082, China}

\author{Li Zhou }
\affiliation{Chinese Academy of Sciences South America Center for Astronomy, National Astronomical Observatories, Chinese Academy of Sciences, Beijing 100101, China; {\it lizhou@bao.ac.cn}}

\author{Bo Ma}
\affil{School of Physics and Astronomy, Sun Yat-sen University, Zhuhai 519082, China; {\it mabo8@mail.sysu.edu.cn}}
\affil{CSST Science Center for the Guangdong-Hong Kong-Macau Great Bay Area, Sun Yat-sen University, Zhuhai 519082, China}

\nocollaboration
%%%%%%%%%%%%%%%%%%%%%%%%%%%%%%%%%%%%%%正文%%%%%%%%%%%%%%%%%%%%%%%%%%%%%%%%%%%%%%%%%%
\begin{abstract}

% TEMP offers a powerful tool for us to achieve several goals. First, we can refine the orbital and physical parameters of the known transiting hot Jupiters discovered with ground-based photo- metric surveys. Second, we can identify statistically significant TTVs, which can be caused not only by planet–planet interactions  (Wang et al. 2017), but also by tidal dissipation, as well as apsidal precession caused by the stellar quadrupole moment, general relativity, and long-period planetary/stellar companions  (Pál & Kocsis 2008).

%\begin{abstract}
The ExoEcho project is designed to study the photodynamics of exoplanets by leveraging high-precision transit timing data from ground- and space-based telescopes. 
Some exoplanets are experiencing orbital decay, and transit timing variation  (TTV) is a useful technique to study their orbital period variations. 
In this study, we have obtained transit middle-time data from the Hubble Space Telescope  (HST) observations for 37 short-period exoplanets, most of which are hot Jupiters. 
To search for potential long- and short-term orbital period variations within the sample, we conduct TTV model fitting using both linear and quadratic ephemeris models. 
Our analysis identifies two hot Jupiters experiencing strong periodic decays. 
Given the old age of the host stars of the hot Jupiter population, our findings call for a scenario where HJs are continuously being destructed and created. 
%that the formation of hot Jupiters may still be an active process. 
Our study demonstrates the importance of incorporating high-precision transit timing data to TTV study in the future. 

\end{abstract}
%tidal orbital decays, gravitational interactions with other bodies, or apsidal precession  (e.g., Agol et al. 2005; Maciejewski et al. 2016a; Bouma et al. 2019).
\keywords{Exoplanet systems --- 
Transit photometry --- Transit timing variation method}

\section{Introduction}
% Xie：
% 云台顾盛宏老师组的孙磊磊有不少TTV的工作应该引用一下。
% https://ui.adsabs.harvard.edu/abs/2017AJ....153...28S
% https://ui.adsabs.harvard.edu/abs/2023MNRAS.520.1642S

% 台湾姜英贵老师组也有一些TTV文章应该引用一下。
% https://arxiv.org/abs/2310.08953
A single transiting planet generally orbits its host star on a Keplerian orbit with a constant orbital period. But recently, studying of the transit timing variations  (TTV), where transits no longer appear at a fixed interval, have become an important topic in the exoplanet research field. Significant long- and short-term TTV can be induced by additional bodies in the planetary systems, the tidal interaction between the host star and the planets, planetary mass loss, apsidal precession, line-of-sight acceleration, and the Applegate mechanism \citep{Applegate92, agol2005-detecting, Holman2005, Lai12,  Bailey19}.
%would exist which means that transits no longer appear at a fixed interval, would exist in principle if there are additional bodies in the planetary systems.
TTV analysis not only facilitates the detection of additional planetary companions, but also can be used to characterize orbital resonances and provide insights into the internal structure and composition of exoplanets \citep{agol2005-detecting, Holman2005, Xie2013, Nesvorny2008}. Thus, it has become an important tool in exoplanetary research field, revealing key properties of planetary systems that might otherwise remain undetected \citep{Jontof2015, deck2015-measurement, Wang18, Wang21, Ivshina22, Kokori22, Wang24}.

Measuring orbital period variations in exoplanets through the TTV measurements can enhance our understanding of their formation theory, as the current orbital dynamic properties of a system retains information about its past evolution history. % orbital properties
%the investigation of TTVs also provides valuable constraints on planet formation theory.
%boma：这里马老师需要再添加一些科学介绍背景，ttv的过去的成果。
\citet{Patra2020} have searched evidence for tidal orbital decay using transit-timing data of twelve hot Jupiters. 
\citet{Yee20} found secular decay in the orbital period of WASP-12~b using TTV measurements,  which they attribute to the tidal interaction between the planet and its hot star \citep[see also][]{Turner21, Wang24}. 
By analyzing 28~yrs of TTV observation, \citet{Maciejewski20} have put a constraint on the tidal quality factor of the host star of WASP-18~b. \citet{Davoudi21} gave a lower limit on the tidal quality factor of WASP-43 by measuring the mid-transit times of WASP-43~b. 
TTV studies also have significant implications for planning future follow-up transit observations, where an accurate transit window prediction is needed \citet{Kokori22}. For example, \citet{Baker20} have provided predictions for shifts in transit times due to tidally driven orbital decay of exoplanets.
%It is also can be used to investigate the formation mechanism of short period exoplanets and brown dwarfs, where in some formation scenarios, tidal damping of orbits are needed. 

%还需要补充些参考文献，可以看下师姐论文，包括王松虎老师他们的TEMP project，也有好几篇。
%xinyue:加了sun2017，a-thano姜英贵老师组2023文章, 松虎老师他们的TEMP2021
% Previous important studies of TTV include the works of and more recently. 
Most of the past exoplanet TTV studies have utilized data from various sources, including space-based telescopes and ground-based observatories \citep{Mazeh2013, Holczer2016, Hadden2017,sun2017-refined, Kane2019, Ivshina22,a-thano2023-revisiting, Wang21, Wang24}. 
Despite significant advancements in the TTV research field, many of these studies are limited by their reliance on archival data from diverse observational projects, a substantial portion of which is ground-based and thus susceptible to many unknown systematics \citep{Mallonn19, Ivshina22, Alvarado24, Wang24}.

In this paper, we propose the incorporation of data from the Hubble Space Telescope  (HST) into the study of TTVs of exoplanets, as part of our ExoEcho project. 
The ExoEcho  (Exoplanet Ephemerides CHange Observation) project is designed to study the photodynamics of exoplanets by leveraging high-precision transit timing data from ground- and space-based telescopes. In our previous two papers \citep{Wang24, Zhang24}, we have studied the TTV behaviors of hot Jupiters mostly using the TESS \citep{Ricker2015} mission data.
By integrating the high precision and reliable HST timing data into the TTV modeling, we expect to put more precise constraints on the TTV properties of exoplanets and identify new candidates exhibiting signs of long- and short-term orbital period variations. 
% The addition of HST observations will enhance the sensitivity and precision of TTV measurements, leading to improved characterization of exoplanet orbital period variations.

The paper is organized as follows: In Section 2, we introduce the transit timing data used in this study. Section 3 presents our analysis methods, including the algorithms and models applied to extract and interpret the TTV signals. In Section 4, we present the TTV results for all thirty-seven exoplanet systems in our sample. A detailed discussion of our findings and their implications for planetary system formation and evolution is presented in Section 5. Finally, we conclude this study with a summary in Section 6.

\section{Sample Selection and Data Analysis}
% %describe where did you get the data, and properties of the data.
\subsection{Sample Selection}
% Here, we selected short orbital period transiting exoplanets observed by the Hubble Space Telescope  (HST) and the Transiting Exoplanet Survey Satellite  (TESS). 
% Additionally, all the planets should have early transiting data, mainly from the first research paper, which provides an important longer-time baseline for analyzing long-term orbital period variations. 
Here we select short orbital period transiting exoplanets from \citet{Wang24}, which have early transiting data and observed by the HST and the TESS, providing an important longer-time baseline for analyzing long-term orbital period variations.
In the end, our sample consists of 37 exoplanets, ranging from warm Neptunes to hot Jupiters. 
We summarize the stellar and planetary parameters for these systems in Table~\ref{tab:parameters}.
All the selected exoplanets have an orbital period $\lesssim 10$~days. 

%The planets included in the sample are HAT-P-2 b, HAT-P-3 b, HAT-P-11 b, HAT-P-12 b, HAT-P-17 b, HAT-P-18 b, HAT-P-24 b, HAT-P-26 b, HAT-P-32 b, HAT-P-38 b, HAT-P-41 b, HD~97658 b, KELT-7 b, KELT-11 b, TrES-4 b, WASP-4 b, WASP-6 b, WASP-12 b, WASP-17 b, WASP-18 b, WASP-19 b, WASP-29 b, WASP-39 b, WASP-43 b, WASP-62 b, WASP-63 b, WASP-69 b, WASP-76 b, WASP-79 b, WASP-80 b, WASP-96 b, WASP-98 b, WASP-101 b, WASP-117 b, WASP-121 b, WASP-127 b and WASP-178 b.

%插入参数表格,表头参考Tsiaras和liz，确定参数来源文献放最后一列
\setlength{\textfloatsep}{1pt}
\begin{deluxetable*}{lccccccccccc}[ht]
\tablecaption{Stellar and Planetary Parameters of Selected Exoplanet Systems \label{tab:parameters}}
\tablewidth{0pt}
\tabletypesize{\scriptsize}
% \tabletypesize{\footnotesize} 
\tablehead{
  \colhead{Planet} & \colhead{[Fe/H]$_\star$} & \colhead{$T_\star$} & \colhead{log ($g_\star$)} & \colhead{$M_p$} & \colhead{$R_p/R_\star$} & \colhead{$P$} & \colhead{$i$} & \colhead{$a/R_\star$} & \colhead{$e$} & \colhead{$\omega$} & \colhead{Ref.\tablenotemark{*}}\\
  % \vspace{-1cm}
  \colhead{} & \colhead{} & \colhead{ (K)} & \colhead{ (cgs)} & \colhead{ ($M_\mathrm{\rm Jup}$)} & \colhead{} & \colhead{ (days)} & \colhead{ (deg)} & \colhead{} & \colhead{} & \colhead{ (deg)} & \colhead{} \\
  % \vspace{-1cm}
}
\startdata
HAT-P-2 b~& 0.12 & 6290 & 4.2 & 9.04 & 0.07227 & 5.63346785 & 86.72 & 8.99 & 0.5171 & 185.22 & 1\tablenotemark{a,b}; 2\tablenotemark{a,b}; 3\tablenotemark{a}; 4\tablenotemark{b}; 5\tablenotemark{b}; 6\tablenotemark{b}; 7\tablenotemark{b}\\
HAT-P-3~b& 0.27 & 5185 & 4.56 & 0.596 & 0.1063 & 2.89973815 & 87.1 & 10.4 & \ldots & \ldots & 8\tablenotemark{a,b}; 3\tablenotemark{a}; 9\tablenotemark{a}; 6\tablenotemark{b}; 10\tablenotemark{b}; 14\tablenotemark{b} \\
HAT-P-11~b& 0.31 & 4780 & 4.6 & 0.081 & 0.0576 & 4.88780201 & 88.5 & 15.58 & 0.2 & 355.2 & 11\tablenotemark{a,b}; 5\tablenotemark{b}; 12\tablenotemark{b}\\
HAT-P-12~b& -0.29 & 4650 & 4.61 & 0.211 & 0.1406 & 3.21305762 & 89 & 11.77  & \ldots & \ldots & 13\tablenotemark{a,b}; 5\tablenotemark{b}; 6\tablenotemark{b}; 10\tablenotemark{b}; 14\tablenotemark{b}\\
HAT-P-17 b& 0 & 5246 & 4.52 & 0.534 & 0.1238 & 10.33853522 & 89.2 & 22.6 & 0.342 & 201 & 15\tablenotemark{a,b}; 5\tablenotemark{b}; 6\tablenotemark{b}\\
HAT-P-18~b& 0.1 & 4803 & 4.57 & 0.197 & 0.1356 & 5.50802941 & 88.53 & 16.39 & \ldots & \ldots & 16\tablenotemark{a,b}; 17\tablenotemark{a}; 5\tablenotemark{b}; 6\tablenotemark{b}; 18\tablenotemark{b}\\
HAT-P-24~b & -0.16 & 6373 & 4.27 & 0.685 & 0.097 & 3.35524439 & 88.6 & 7.6 & 0.067 & 197 & 19\tablenotemark{a,b}; 5\tablenotemark{b}; 6\tablenotemark{b}\\
HAT-P-26~b& -0.04 & 5079 & 4.56 & 0.059 & 0.0737 & 4.2345002 & 88.6 & 13.1 & 0.12 & 54 & 20\tablenotemark{a,b}; 5\tablenotemark{b}\\
HAT-P-32~b& -0.04 & 6207 & 4.329 & 0.68 & 0.1489 & 2.150008197 & 89 & 5.34 & 0.16 & 50 & 21\tablenotemark{a,b}; 22\tablenotemark{a,b}; 5\tablenotemark{b}; 6\tablenotemark{b}\\
HAT-P-38~b& 0.06 & 5330 & 4.46 & 0.267 & 0.0918 & 4.64032787 & 88.3 & 12.2 & 0.07 & 240 & 23\tablenotemark{a,b}; 6\tablenotemark{b} \\
HAT-P-41~b& 0.21 & 6390 & 4.14 & 0.812 & 0.1028 & 2.69404968 & 87.7 & 5.44 & \ldots & \ldots & 24\tablenotemark{a,b}; 6\tablenotemark{b}\\
HD~97658~b& -0.23 & 5170 & 4.63 & 0.02375 & 0.0311 & 9.4893037 & 89.8 & 26.2 & \ldots & \ldots & 25\tablenotemark{a,b}; 26\tablenotemark{a}; 27\tablenotemark{a}; 28\tablenotemark{b}; 29\tablenotemark{b}\\
KELT-7~b& 0.14 & 6789 & 4.15 & 1.28 & 0.091 & 2.7347656 & 83.8 & 5.53 & \ldots & \ldots & 30\tablenotemark{a,b} \\
KELT-11~b& 0.18 & 5370 & 3.73 & 0.195 & 0.051 & 4.7362006 & 85.3 & 4.98 & 0.0007 & 359 & 31\tablenotemark{a,b}; 32\tablenotemark{b} \\
TrES-4~b& 0.14 & 6200 & 4.06 & 0.84 & 0.1045 & 3.55392889 & 83.1 & 6.14 & \ldots & \ldots & 33\tablenotemark{a,b}; 5\tablenotemark{b}; 6\tablenotemark{b}; 34\tablenotemark{b}\\
WASP-4~b& 0 & 5500 & 4.3 & 1.186 & 0.152 & 1.338231388 & 89.1 & 5.451 & \ldots & \ldots & 35\tablenotemark{a,b}; 36\tablenotemark{a,b}; 5\tablenotemark{b}; 6\tablenotemark{b}; 37\tablenotemark{b}; \\ 
WASP-6~b & \ldots & 5450 & 4.5 & 0.503 & 0.1446 & 3.36100215 & 88.5 & 10.9 & 0.054 & 1.7 & 38\tablenotemark{a,b}; 6\tablenotemark{b}\\
WASP-12~b& 0.3 & 6300 & 4.18 & 1.47 & 0.1178 & 1.091419179 & 83.4 & 3.04 & \ldots & \ldots & 39\tablenotemark{a,b}; 40\tablenotemark{a,b}; 5\tablenotemark{b}; 6\tablenotemark{b}\\
WASP-17~b& -0.25 & 6550 & 4.24 & 0.486 & 0.1235 & 3.73548545 & 87.1 & 7.03 & \ldots & \ldots & 41\tablenotemark{a,b}; 42\tablenotemark{a,b}; 43\tablenotemark{a}; 5\tablenotemark{b}; 6\tablenotemark{b}; 37\tablenotemark{b}\\
WASP-18~b& 0 & 6400 & 4.4 & 10.3 & 0.09716 & 0.941452417 & 84.9 & 3.562 & 0.0091 & 269 & 44\tablenotemark{a,b}; 45\tablenotemark{a}; 5\tablenotemark{b}; 6\tablenotemark{b}; 37\tablenotemark{b}\\
WASP-19~b& 0.2 & 5500 & 4.5 & 1.069 & 0.1409 & 0.788839092 & 78.8 & 3.46 & 0.002 & 259 & 46\tablenotemark{a,b}; 47\tablenotemark{a}; 5\tablenotemark{b}; 6\tablenotemark{b}; 48\tablenotemark{b}\\
WASP-29~b& 0.11 & 4800 & 4.54 & 0.244 & 0.0982 & 3.92271183 & 89.2 & 12.36 & \ldots & \ldots & 49\tablenotemark{a,b}; 50\tablenotemark{a}; 6\tablenotemark{b};\\
WASP-39~b& -0.12 & 5400 & 4.498 & 0.283 & 0.1457 & 4.05528043 & 87.75 & 11.37 & \ldots & \ldots & 51\tablenotemark{a,b}; 52\tablenotemark{a}; 6\tablenotemark{b}; 10\tablenotemark{b}\\
WASP-43~b& -0.05 & 4400 & 4.64 & 1.78 & 0.1594 & 0.813474056 & 82.11 & 4.867 & \ldots & \ldots & 53\tablenotemark{a,b}; 54\tablenotemark{a}; 6\tablenotemark{b}; 55\tablenotemark{b};\\
WASP-62~b& 0.04 & 6280 & 4.32 & 0.57 & 0.1109 & 4.41193868 & 88.3 & 9.52 & \ldots & \ldots & 56\tablenotemark{a,b}; 6\tablenotemark{b}; 57\tablenotemark{b}\\
WASP-63~b& 0.08 & 5570 & 4.01 & 0.38 & 0.078 & 4.37808205 & 87.8 & 6.49 & \ldots & \ldots & 56\tablenotemark{a,b}; 6\tablenotemark{b} \\
WASP-69~b& 0.14 & 4715 & 4.54 & 0.26 & 0.1336 & 3.86813888 & 86.71 & 11.953 & \ldots & \ldots & 58\tablenotemark{a,b}; 6\tablenotemark{b}; 59\tablenotemark{b}\\
WASP-76~b& 0.23 & 6250 & 4.13 & 0.92 & 0.109 & 1.80988043 & 88 & 4.07 & \ldots & \ldots & 60\tablenotemark{a,b}; 61\tablenotemark{b}\\
WASP-79~b& 0.03 & 6600 & 4.06 & 0.9 & 0.113 & 3.66239163 & 85.4 & 7.02 & \ldots & \ldots & 62\tablenotemark{a,b}; 6\tablenotemark{b}; 57\tablenotemark{b}\\
WASP-80~b& -0.14 & 4145 & 4.69 & 0.538 & 0.1714 & 3.06785251 & 89.02 & 12.63 & 0.002 & 94 & 63\tablenotemark{a,b}; 64\tablenotemark{a,b}; 6\tablenotemark{b}\\
WASP-96~b& 0.14 & 5540 & 4.42 & 0.48 & 0.1175 & 3.42525674 & 85.6 & 9.255 & \ldots & \ldots & 65\tablenotemark{a,b}; 6\tablenotemark{b}\\
WASP-98~b& -0.6 & 5525 & 4.58 & 0.922 & 0.1582 & 2.96264191 & 86.38 & 10.92 & \ldots & \ldots & 65\tablenotemark{a,b}; 66\tablenotemark{a}; 6\tablenotemark{b}\\
WASP-101~b& 0.2 & 6400 & 4.34 & 0.5 & 0.1122 & 3.585707 & 85
& 8.445 & \ldots & \ldots & 65\tablenotemark{a,b}; 6\tablenotemark{b}\\
WASP-117~b& -0.11 & 6038 & 4.28 & 0.2755 & 0.09 & 10.0205933 & 89.1 & 17.4 & 0.302 & 242 & 67\tablenotemark{a,b}; 6\tablenotemark{b}\\
WASP-121~b& 0.13 & 6459 & 4.24 & 1.183 & 0.1245 & 1.274924762 & 87.6 & 3.754 & \ldots & \ldots & 68\tablenotemark{a,b}; 69\tablenotemark{b}\\
WASP-127~b& -0.18 & 5620 & 4.18 & 0.18 & 0.1004 & 4.17806513 & 88.2 & 7.95 & \ldots & \ldots & 70\tablenotemark{a,b}; 71\tablenotemark{a}\\
WASP-178~b& 0.21 & 9350 & 4.35 & 1.66 & 0.1115 & 3.3448285 & 85.7 & 7.17 & \ldots & \ldots & 72\tablenotemark{a,b}; 73\tablenotemark{b}\\
\enddata
\tablenotetext{*}{Most of the ephemerides parameters are from \citet{Kokori2023}, except WASP-178 b.}
\tablenotetext{a}{References for stellar and planetary parameters.}
\tablenotetext{b}{We also include references of RV data available for each planet.}
\tablecomments{The transit mid-time and depth are not specified because they are treated as free parameters during fitting. References: 1. \citet{Bakos2007}; 2. \citet{Pal2010}; 3. \citet{Torres2008}; 4. \citet{Loeillet2008}; 5. \citet{Knutson2014Friends}; 6. \citet{Bonomo2017}; 7. \citet{deBeurs23}; 8. \citet{Torres2007}; 9. \citet{Chan2011}; 10. \citet{Mancini2018}; 11. \citet{Bakos2010}; 12. \citet{Yee2018} 13. \citet{Hartman2009}; 14. \citet{Ment2018}; 15. \citet{Howard2012}; 16. \citet{Hartman2011_18_19}; 17. \citet{Kirk2017}; 18. \citet{Esposito2014}; 19. \citet{Kipping2010}; 20. \citet{Hartman2011_26}; 21. \citet{Hartman2011_32_33}; 22. \citet{Wang2019}; 23. \citet{Sato2012}; 24. \citet{Hartman2012}; 25. \citet{Howard2011}; 26. \citet{Knutson2014}; 27. \citet{Van2014}; 28. \citet{Dragomir2013}; 29. \citet{Rosenthal2021}; 30. \citet{Bieryla2015}; 31. \citet{Pepper2017}; 32. \citet{Beatty2017}; 33. \citet{Mandushev2007}; 34. \citet{Sozzetti2015}; 35. \citet{Wilson2008}; 36. \citet{Bouma2019}; 37. \citet{Triaud2010}; 38. \citet{gillon_discovery_2009}; 39. \citet{Hebb2009}; 40. \citet{Collins2017}; 41. \citet{Anderson2010}; 42. \citet{Anderson2011}; 43. \citet{Sedaghati_Potassium_2016}; 44. \citet{Hellier2009}; 45. \citet{Shporer2019}; 46. \citet{Hebb2010}; 47. \citet{Wong2016}; 48. \citet{Hellier2011}; 49. \citet{Hellier2010}; 50. \citet{Gibson2013}; 51. \citet{Faedi_wasp39_2011}; 52. \citet{Maciejewski2016_30_37}; 53. \citet{Hellier_closest_2011}; 54. \citet{Hoyer2016}; 55. \citet{Esposito2017}; 56. \citet{Hellier2012} 57. \citet{Brown2017}; 58. \citet{Anderson2014}; 59. \citet{Casasayas-Barris2017}; 60. \citet{West_Three_2016}; 61. \citet{Ehrenreich2020}; 62. \citet{Smalley_WASP7879_2012}; 63. \citet{Triaud2013}; 64. \citet{Triaud2015}; 65. \citet{Hellier2014}; 66. \citet{Mancini2016}; 67. \citet{Lendl2014}; 68. \citet{Delrez2016}; 69. \citet{Bourrier2020}; 70. \citet{Lam2017}; 71. \citet{Palle2017}; 72. \citet{Hellier2019}; 73. \citet{Rodriguez2020}
}
\end{deluxetable*}

\subsection{Hubble Data} 
%描述下哈勃观测情况，以及如何得到光变曲线的。
% 发现个别有些行星的数据偏差太大，尝试寻找解决方案，虽然涉及到的一般都是常数周期的系统。
For the 37 exoplanets in our sample, we download their HST/WFC3 slitless spectroscopy transit observation data \citep{doi:10.17909/t97p46} with the IR G102 or G141 grism from the NASA Mikulski Archive for Space Telescopes archive. 
All observations are executed in the spatial scanning mode, which can maximize the signal-to-noise ratio and avoid saturation for these bright targets.
%We present the analysis of 37 short-period exoplanets observed with the HST/WFC3 camera, in the spatial scanning mode. 
%Data were obtained from the publicly accessible pages of the NASA Mikulski Archive for Space Telescopes archive. 
%In our data reduction, we exclude 
%The first HST orbit was excluded from the analyzed transit observations due to significant systematic effects. 
A complete list containing the observation proposal information, the number of transits covered, and the HST orbits spent for the observation are shown in Table~\ref{tab:Proposal Information}.
Most of our targets have one to five transits covered by the HST observations.

\begin{deluxetable}{ccccc} [ht]
\tablecaption{HST Observations Analyzed \label{tab:Proposal Information}}
% \centering
%\tablecolumns{6}
%\tablenum{2}
\tabletypesize{\scriptsize} % Adjust table font size
\tablewidth{0pt}
\tablehead{
\colhead{} & \colhead{} & \colhead{} & \colhead{} & \colhead{HST}\\
% \vspace{-3cm}
\colhead{} & \colhead{} & \colhead{} & \colhead{Transits} & \colhead{Orbits}\\
% \vspace{-3cm}
\colhead{Planet} & \colhead{Proposal ID} & \colhead{Proposal PI} & \colhead{Covered} & \colhead{Used}
}
\startdata
HAT-P-2 b & 16194 & Jean-Michel Desert & 1 & 5 \\
HAT-P-3 b & 14260 & Drake Deming & 2 & 8 \\
HAT-P-11 b & 12449 & Drake Deming & 1 & 3\\
HAT-P-11 b & 14793 & Jacob Bean & 5 & 15\\
HAT-P-12 b & 14260 & Drake Deming & 2 & 8\\
HAT-P-17 b & 12956 & Catherine Huitson & 1 & 4\\
HAT-P-18 b & 14099 & Thomas Evans-Soma & 1 & 3 \\
HAT-P-18 b & 14260 & Drake Deming & 2 & 8\\
HAT-P-24 b & 16736 & David Sing & 1 & 4 \\
HAT-P-24 b & 16587 & Frederick Dauphin & \ldots & \ldots \\
HAT-P-26 b & 14110 & David Sing & 1 & 4 \\
HAT-P-26 b & 14260 & Drake Deming & 2 & 8 \\
HAT-P-32 b & 14260 & Drake Deming & 1 & 4 \\
HAT-P-38 b & 14260 & Drake Deming & 2 & 8 \\
HAT-P-41 b & 14767 & David Sing & 1 & 4 \\
HD 97658 b & 13501 & Heather Knutson & 1 & 4 \\
HD 97658 b & 13665 & Bjorn Benneke & 2 & 8 \\
KELT-7 b & 14767 & David Sing & 1 & 4 \\
KELT-11 b & 15926 & Knicole Colon & 1 & 8 \\
KELT-11 b & 15255 & Knicole Colon & 1 & 7 \\
TrES-4 b & 12181 & Drake Deming & 1 & 4 \\
WASP-4 b & 12181 & Drake Deming & 1 & 4 \\
WASP-6 b & 14767 & David Sing & 1 & 4 \\
WASP-12 b & 13467 & Jacob Bean & 4 & 16 \\
WASP-12 b & 12330 & Mark Swain & 1 & 4 \\
WASP-12 b & 16236 & Taylor Bell & 1 & 4 \\
WASP-17 b & 14918 & Hannah Wakeford & 3 & 12\\
WASP-17 b & 12181 & Drake Deming & 1 & 4 \\
WASP-18 b & 12181 & Drake Deming & 1 & 4 \\
WASP-19 b & 12181 & Drake Deming & 1 & 4 \\
WASP-29 b & 14260 & Drake Deming & 1 & 4 \\
WASP-39 b & 14260 & Drake Deming & 1 & 4 \\
WASP-43 b & 13467 & Jacob Bean & 4 & 16 \\
WASP-62 b & 14767 & David Sing & 1 & 4 \\
WASP-63 b & 14642 & Kevin Stevenson & 1 & 7 \\
WASP-69 b & 14260 & Drake Deming & 1 & 3 \\
WASP-76 b & 14767 & David Sing & 1 & 4 \\
WASP-79 b & 14767 & David Sing & 1 & 4 \\\
WASP-80 b & 14260 & Drake Deming & 1 & 3 \\
WASP-96 b & 15469 & Nikolay Nikolov & 1 & 4 \\
WASP-98 b & 16736 & David Sing & 1 & 3 \\
WASP-101 b & 14767 & David Sing & 1 & 4 \\
WASP-117 b & 15301 & Ludmila Carone & 1 & 10 \\
WASP-121 b & 15134 & Thomas Evans-Soma & 1 & 4 \\
WASP-121 b & 14468 & Thomas Evans-Soma & 1 & 4 \\
WASP-127 b & 14619 & Jessica Spake & 1 & 4 \\
WASP-178 b & 16450 & Joshua Lothringer & 2 & 8 \\
\enddata
\tablecomments{The HAT-P-24's proposal 16587 provides the flt.fits file for proposal 16736.}
\end{deluxetable}

\begin{figure}[h!]
\plotone{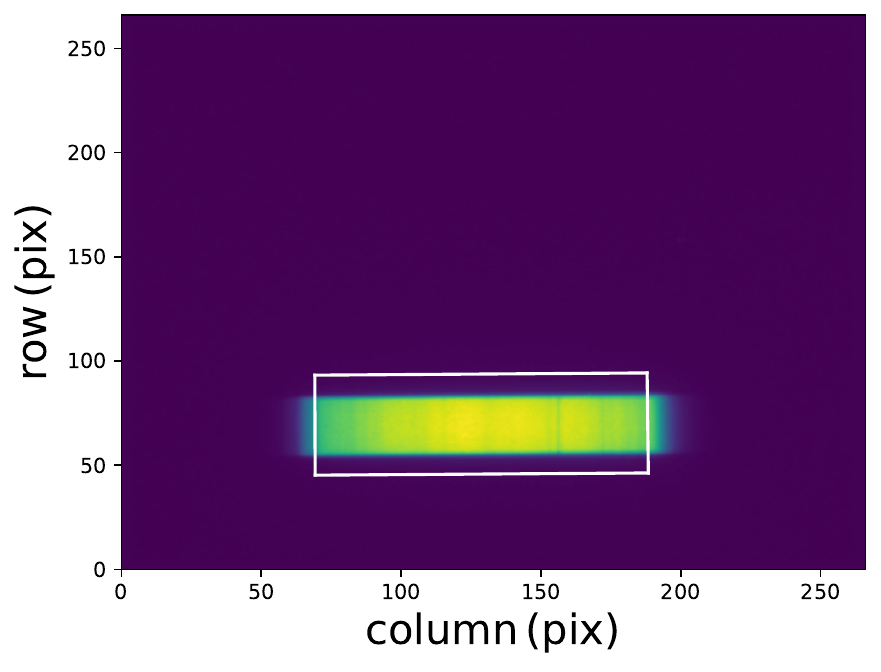}
\caption{The detector image of a forward scan
\label{fig:forward-scan}}
\end{figure}

\begin{figure}[h!]
\plotone{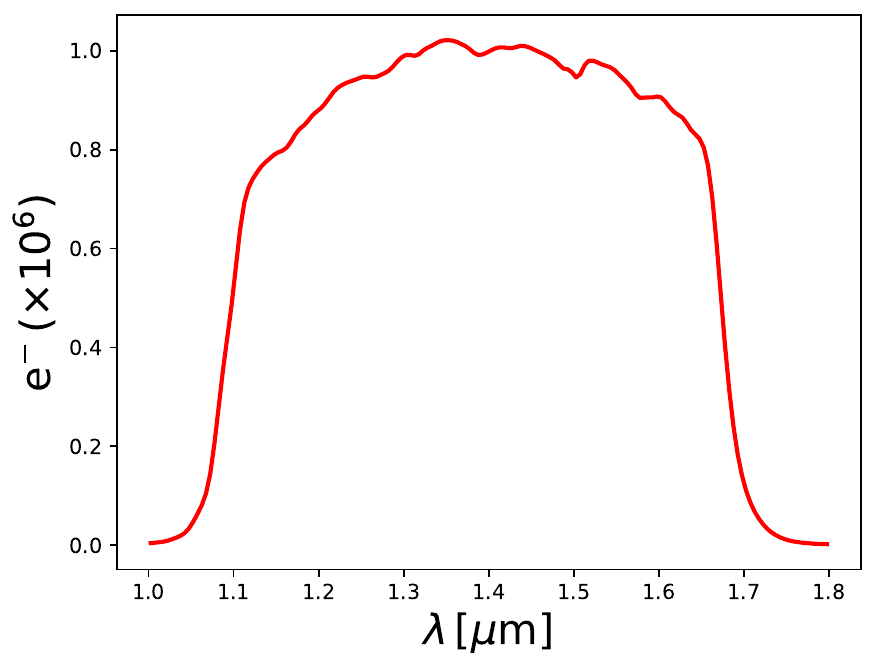}
\caption{Extracted 1D spectrum using Iraclis 
\label{fig:1D-spectrum}}
\end{figure}

\begin{figure}[ht]
\centering
\includegraphics[width=0.45\textwidth]{ 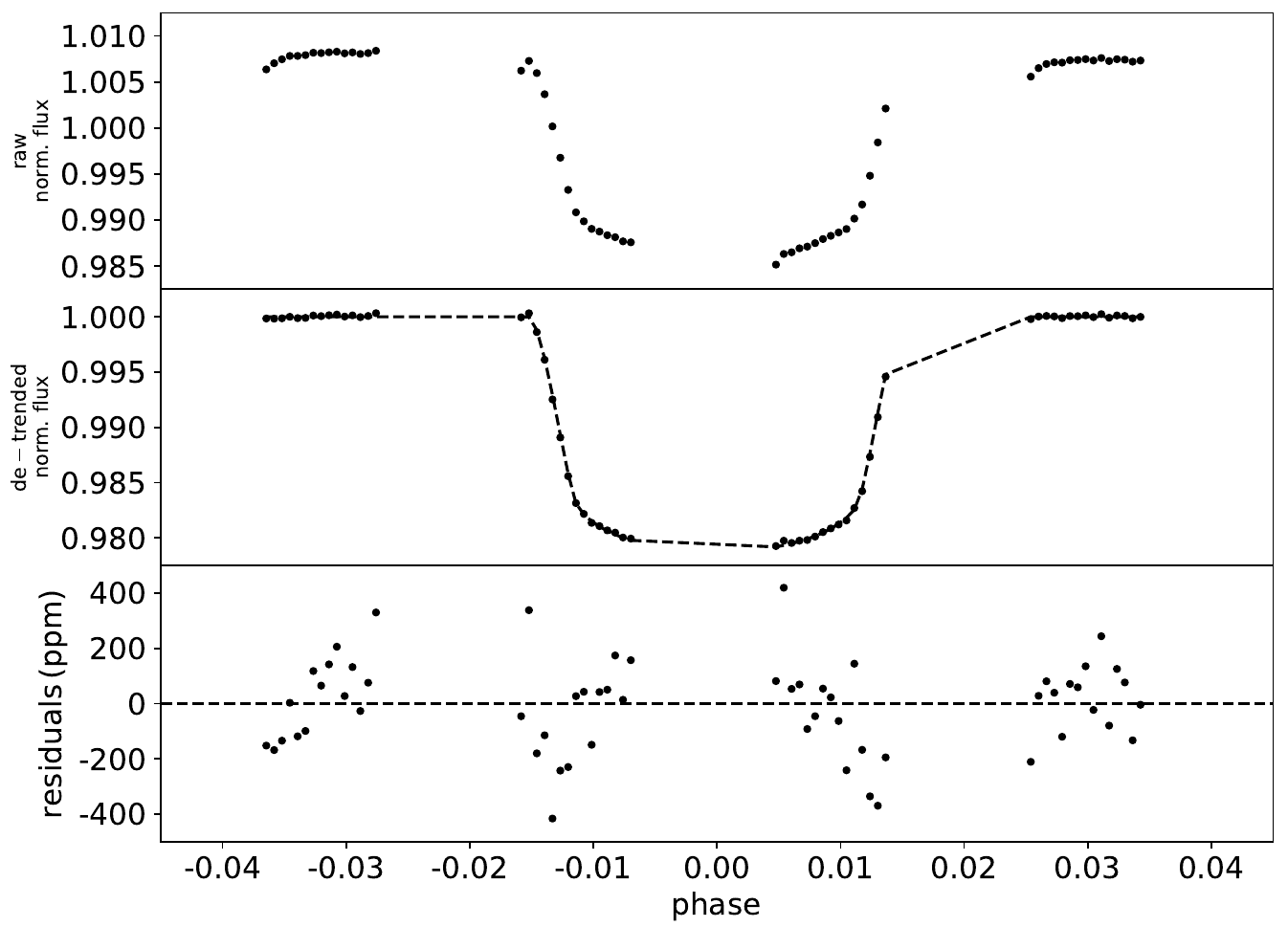}
% \plotone{ HAT-P-12b_2015_white_fitting.pdf}
\caption{Top: the normalized raw white light curve of HAT-P-12 from the 2015 observation from HST/WFC3  (PI: Drake Deming, Proposal ID: 14260). Middle: the same raw white light curve divided by the best-fit model of the instrument systematics. Bottom: residuals from the best-fit transit model light curve. 
\label{fig:white-fitting}}
\end{figure}

The data reduction for transit observation from HST/WFC3 is the same as in our previous studies \citep{Zhou22, Zhou23}.
We start our data reduction from the raw HST/WFC3 spatially scanned spectroscopic images using \href{https://github.com/ucl-exoplanets/Iraclis}{Iraclis} \citep{Tsiaras2016, Tsiaras2016_55, Tsiaras2018}. 
%for the analysis of spatially scanned spectroscopic images from HST/WFC3, 
The basic calibration includes the following steps: zero-read subtraction, reference pixels correction, non-linearity correction, dark current subtraction, gain conversation, sky background subtraction, calibration, flat-field correction, and bad pixels/cosmic rays correction. 
Following the basic calibration, 1D spectra are extracted using the standard optimal extraction technique \citep{opt1, opt2, opt3, opt4, opt5}.
An example of the extraction region and the corresponding extracted 1D spectrum are presented in Figures~\ref{fig:forward-scan} and ~\ref{fig:1D-spectrum}. 
%We considered one broadband  (white) covering the whole wavelength range where the G141 grism is sensitive  (1.088-1.68$\mu m$). 
%xinyue: 除了G141  (1075-1700nm)还用了G102  (800-1150nm),换成了官网查到的波段数据。
We sum the G141 spectrum from 1.1 to 1.7~$\mu$m or the G102 spectrum from 0.8 to 1.15~$\mu$m to obtain raw white light curves of transiting exoplanets from each exposure, as shown in the top panel of Figure~\ref{fig:white-fitting}.
%\textcolor{blue}{We summed the G141 spectrum from 1.1 to 1.7~$\mu$m to obtain the raw white light curve from observation using G141 prism. Similarly, we summed the G102 spectrum from 0.8 to 1.15~$\mu$m for the raw white light curve from G102.}
%We summed the G141 spectrum from 1.088 to 1.68~$\mu$m for each exposure to obtain the raw white light curve of the planetary system, as shown in the top panel of Figure~\ref{fig:white-fitting}. 
Outliers are then removed from the final raw white light curves.
%We then summed the G141 spectrum from 1.088-1.68$\mu m$ of each exposure to obtain the raw white light curve for planetary system, as shown in the top panel of Figure~\ref{fig:white-fitting}. Outliers were also removed from the final raw white light curves.
%boma 一般在 Figure后面加一个~ ,保持它们和数字的链接。
%mxy Table后面也加~么，还有那些行星 eg：WASP-12~b?

%All raw white light curves were fitted for the $R_p/R_\star$ and $T_0$ parameters, during which we used fixed values for the stellar, planetary, and orbital parameters from literature  (see Table~\ref{tab:parameters}). 
We obtain the middle transit time parameter $T_{\rm mid}$ for each transit observation by fitting all the raw white light curves with the transit model from \citet{Mandel2002}. 
During the fitting process, we fix the stellar, planetary, and orbital parameters to values from literature  (see Table~\ref{tab:parameters}) and allow only $T_0$ and $R_p/R_\star$ to vary as free parameters. 
We model the stellar limb-darkening effect using the nonlinear formula proposed by \citet{Claret2000}. The limb-darkening coefficients are derived from specific intensity profiles evaluated at 100 angles, calculated directly from the ATLAS model \citep{Howarth2011}, as all our target stars have effective temperatures above 4,000~K  (see Table~\ref{tab:parameters}). 
%xinyue:老师这里有点疑问
To correct for time-dependent systematics in the HST/WFC3 observations \citep{Kreidberg2015,Wakeford2016,Evans2016,Line16,Wakeford2017_101,Tsiaras2018}, we fit the transit model together with a normalization factor $n_w$ and an instrumental systematics function $R (t)$, following the studies of \citet{Kreidberg2014} and \citet{Tsiaras2016,Tsiaras2018}. 
% To correct the time-dependent systematics from the observation of WFC3/HST \citep{Kreidberg2015,Evans2016,Line2016,Wakeford2017_101}, we fit the transit model together with a normalization factor $n_w$ and an instrumental systematics function R (t), following the studies of \citet{Kreidberg2014} and \citet{Tsiaras2016}. 
%Figure~\ref{fig:white-fitting} shows an example of the fitting results. 
%In Figure~\ref{fig:correlations}, we show an example corner plot of transit model fitting for the raw HST/WFC3 light curve of the HAT-P~12 system.
%We can also see an example of correlations in Figure~\ref{fig:correlations}. 
% In the observations for HAT-P-12~b  (14260), HAT-P-18 b  (14260), WASP-4 b  (12181), WASP-17 b  (14918), WASP-39 b  (14260), and WASP-76 b  (14767), we excluded one or two defect points with significant deviations from the overall data. 
%Finally, except for some eclipses, we obtained 64 HST light curves and mid-transit times, which are summarized in Table~\ref{tab:times}.

To double check the fitting result of the Iraclis package, we also fit the raw white light curves using the \href{https://github.com/hpparvi/PyTransit}{PyTransit} \citep{Parviainen2015} and \href{https://github.com/lkreidberg/batman}{batman} \citep{Kreidberg2015_batman} packages. 
We adopt the fitting result from the PyTransit fitting when we deem it feasible.
In the end, we obtain a total of 64 mid-transit times for the 37 planetary systems in our sample, which are summarized in Table~\ref{tab:times}.

% 图纵坐标文字不清晰
% \textcolor{red}{最下面的res ACF 也许可以从图里删除，完全没有描述。}
% \begin{figure}[ht]
% \plotone{ HAT-P-12b_2015_white_correlations.pdf}
% \caption{ Corner plot showing the results from the transit model fitting for the raw HST/WFC3 light curve of the HAT-P-12 system.
% %The white fitting correlations. 
% % \textcolor{red}{Xinyue, please check and make sure that figures 1-4 are taken from your own data processing for this work.} \textbf{OK, these figures are derived from my data processing of the HAT-P 12 system using Iraclis.}
% \label{fig:correlations}}
% \end{figure}

% \setlength{\textfloatsep}{0pt}
\begin{deluxetable}{ccccc}[ht]
\tablecaption{Middle transit times extracted by fitting the raw white light curves from the HST/WFC3 observation. \label{tab:times}}
\tablewidth{0pt}
\tabletypesize{\scriptsize}
% \tabletypesize{\footnotesize} 
\tablehead{
  \colhead{Planet} & \colhead{Proposal ID} & \colhead{$T_{\rm mid}  (\mathrm{BJD_{TDB}})$} & \colhead{Uncertainty  (days)} & \colhead{Tools} 
}
\startdata
HAT-P-2 b& 16194 & 2459204.10902 & 0.0002345 & I\tablenotemark{*}
\enddata
\tablenotetext{*}{Iraclis  (I) or Iraclis+PyTransit  (IP)}
\tablecomments{Table~\ref{tab:times} is fully published in a machine-readable format, with a sample provided here for reference. 
%\textcolor{red}{The uncertainties in this table are reported as the maximum of the upper and lower uncertainties for consistency.}
}
\end{deluxetable}

\section{Modeling of the Transit Times Data}
In this section, we utilize the middle transit times extracted from the HST/WFC3 data in this study, along with archival transit times from the database of \citet{Wang24} and \citet{Ivshina22}, to study the transit timing variations  (TTVs) of the planets in our sample. For TESS observation, we adopt only results from the study of \citet{Wang24}.

\subsection{Transit-timing Models}
% TTV studies have traditionally adopted two models: the constant period model and the constant period derivative model \citep[eg:][]{Maciejewski2016,Patra2017,Wang24}. %zzx
TTV studies utilize two popular models: the constant period model and the constant period derivative model \citep[eg:][]{Maciejewski2016, Patra2017, Wang24}.
% We fitted these two models to the timing data using the MCMC method. %zzx
We use the MCMC method to fit these models to the data of the transit middle times.

The first model assumes a linear ephemeris with a constant orbital period $P$:
%The first model assumes a linear ephemeris with a constant orbital period P on a circular orbit:
\begin{equation}
T_n = T_0 + nP
\end{equation}
where n is the transit number counting from the zero reference middle transit time $T_0$, and $T_n$ is time of the n-th transit. 
The second model assumes a quadratic ephemeris that the period changes at a constant rate:
\begin{eqnarray}
T_n &=& T_{n-1} + P_{n-1}, \\
P_n &=& P_{n-1} +\frac{P_{n-1}+P_{n}}{2}\dot{P},
\end{eqnarray}
%Wenqin；这里直接用pdot吧
%xinyue: 嗯嗯
% for $\dot{P}$ the period change rate, which is $\frac{dP}{dt} = \frac{1}{P}\frac{dP}{dn}$. This recursion form uses previous values to calculate the next value in a sequence, allowing it to adapt to complex, dynamic changes over time easily. In many physical phenomena, changes are not instantaneous but gradual. %zzx
where the period change rate, denoted by $\dot{P}$, is defined as $\frac{dP}{dt} = \frac{1}{P}\frac{dP}{dn}$. 
This recursive formula utilizes past values to compute subsequent values in a sequence. 
%This recursive formula utilizes past values to compute subsequent values in a sequence, enabling it to adapt to complex, evolving changes over time efficiently. 
% In various physical processes, transitions occur gradually rather than instantaneously.
% % Using the average value better approximates these incremental changes. %zzx
% Using the average value provides a better approximation for these incremental changes.
The three parameters used in the second model are the zero reference transit time $T_0$ corresponding to the zero epoch, the orbital period $P_0$ at $T_0$, and the constant period change rate $\dot{P}$.

% %On the other hand, if the orbit  is slightly eccentric, one may expect the argument of periapse to precess over time due to a number of effects.

\subsection{Transit Timing Data Analysis \label{sec:analysis}}
We use the transit timing data analysis tool \href{https://github.com/AeoN400/PdotQuest}{PdotQuest} \citep{Wang24} to fit the transit middle time for both models. 
In our MCMC analysis \citep{Goodman2010,Foreman-Mackey2013}, we utilize 100 walkers to explore the parameter space, with each walker completing 500 steps. To ensure convergence, we discarded the first 100 steps of each walker as burn-in.
% In our MCMC analysis \citep{Goodman2010,Foreman-Mackey2013}, we employed the following parameter settings: 100 walkers to explore the parameter space, each walker took 500 steps during the sampling process, and the first 100 steps of each walker were burned. 
For all the fitting parameters, we use uniform priors.

To identify and exclude outliers, we apply an iterative fitting method known as `n-$\sigma$ rejection'. 
This method involves calculating the standard deviation of the fitting residuals and removing any data points with residual values exceeding the n-sigma threshold from the residual mean. 
Specifically, we explore the fitting results from within the 5-$\sigma$ and 3-$\sigma$ ranges.

The Bayesian Information Criterion  (BIC) is a statistical tool for model selection among a finite set of competing models \citep{schwarz1978estimating}.
%The Bayesian Information Criterion  (BIC), introduced by \citet{schwarz1978estimating}, is a statistical criterion used for selection among a finite set of models.
% We utilized it to compare the goodness of fit of linear and quadratic models while penalizing for model complexity. %zzx
In this study, we employ BIC to compare fitting results from both of the linear and quadratic models, while accounting for model complexity.
%We used BIC to compare the accuracy of linear and quadratic models while also considering model complexity.
BIC is calculated as $BIC = \chi^2 + klogn$, where k is the number of parameters, and n is the number of observation data points. Lower BIC values suggest better-fitting models and $\Delta BIC > 10$ indicates strong evidence of the model with the lower BIC \citep{doi:10.1080/01621459.1995.10476572}. 

We identify long-term period changes in the samples using two criteria:  (a) the period derivative $\dot{P}$ must be at least 3-$\sigma$ away from zero, indicating a significant deviation from a constant period, and  (b) the difference in Bayesian Information Criterion  ($\Delta BIC = BIC_{linear} - BIC_{quadratic}$) must exceed 10, suggesting strong evidence in favor of the model with the lower BIC. 
%\textcolor{red}{We initially identified xxx candidates showing evidence of long-term period changes.}

In the study of \citet{Ivshina22} and \citet{Wang24}, they have also mentioned one single data point with unrealistically small error bar will produce false positives when looking for orbital period decay using TTV method. %this phenomenon in their papers.
To validate our fitting results, we use the leave-one-out cross-validation  (LOOCV) test following \citet{Wang24}.
% Next, 
% % to verify the credibility of the results, %zzx
% to validate the results, we use the leave-one-out cross-validation \citep[LOOCV;][]{Wang24} test. 
% The test operates by removing one point at a time and refitting the remaining points. %zzx
This test iteratively removes a data point and re-fits the remaining data to assess the impact of one data point on the model fitting.% quadratic model of period change.
%This allows us to identify the data point with the greatest influence and further evaluate the robustness of the period change quadratic model. 
% Besides, we repeated the fitting process due to a general concern about systematic errors and overly optimistic results of the fitting error in light-curve fitting. 
% In detail, we increased the error bars as follows for the data points identified by the LOOCV test that make a heavy difference to the fitting results.  (a) if the data point is from combined light curves and the separate points also exist in the database, we will delete it  (eg: WASP-80 b).  (b) If the error of data points is smaller than 0.0003 days, we will scale it to the smaller value between 0.0003 days and three times its original error. %zz
For some data points identified by the LOOCV test that can significantly affect the fitting results, we also explore different TTV fitting solutions by not simply removing the data point but modifying its error bar in the following way. If the error bar of one data point identified by LOOCV is less than 0.0003~days, we will enlarge it to 0.0003~days or three times its original value, whichever is smaller, and re-do the fitting. 
By conducting this additional test, we can have more robust quadratic model fitting results, which are less impacted by transit time data having unreliably small error bars.
Out of the 37 planets in our sample, three have undergone the additional testing procedure. 
%\textcolor{red}{After the LOOCV test, we now have xxx candidate planets showing strong evidence of long-term period variation.}

% We modified the error bars for specific data points identified by the LOOCV test that significantly impact the fitting results.  (a) If the data point is from combined light curves and also exists as a separate entry in the database, we will remove it  (e.g., WASP-80 b).  (b) If the error of a data point is less than 0.0003 days, we will enlarge it to the smaller value between 0.0003 days and three times its original error.

% For specific cases, we adjusted the error bars of our processed HST data to align with those of the TESS data, amplifying errors in the HST data significantly smaller than the median error in the TESS data. This adjustment aimed to match the median error in the TESS data while preserving the relative error sizes in the original HST data. 
% We used the discovery paper data, with reasonable amplification, and our own obtained data to refit the trend.
%boma v1: 这里估计是referee会仔细问如何amplitify的。我记得汪同学曾经引用过一个timing精度和SNR关系的论文。你可以问一下，我们是否能估算下合理的timing精度。
%xinyue：估算ing
% Then we combined the first point from the discovery paper and our points to refit the trend. 
% We have also performed fits individually for some sources whose processed TESS and HST data are well distributed across epochs. 
% 上面那句话应该是指HST和TESS数据分布的时间跨度比较长而且相对均匀的情况
% Besides, we attempted to amplify the errors of the discovery data that had a significant impact on the LOOCV test and re-fit it again. 

\section{Results \label{sec:result}}
% Xie：前面提到的孙磊磊的文章和姜英贵的文章里，可能也report了一些这里的系统，需要在讨论具体系统的时候引用一下。
% sunleilei组2022的WASP12加了进来, 姜老师组的HAT-P-26的结果也引了进来
In this section, we present the model fitting and testing results for all the planets in our sample. 
After examining the result of each target, we divide the whole sample into four different categories: 6 systems showing signs of long-term period variation, 4 systems showing signs of short-term period variation, 8 systems as `interesting', and 19 systems showing no signs of period variation. The final ephemerides derived in this work are made available online.
We will discuss individual systems from the first three categories next.
All fitting results for the remaining 19 systems belonging to the fourth category are shown in Figure~\ref{fig:all_plots_appendix} in the Appendix.

% We examine all the model fitting and testing results in this section. 
% Using the fitting results and the LOOCV test, 
% In this section, we introduce some systems that probably exhibit long-term period changes using archival, TESS, and HST data, with the initial 2 candidates successfully passing the LOOCV test. 
% We refitted these systems with expanding errors for some heavily influential points. Besides, we also refitted the trend using HST and TESS data. The results show most of the trend can not be considered robust. In conclusion, WASP-4b and WASP-12b exhibit a robust long-term decreasing trend.

% The remaining transiting planets show no significant evidence of period variation, and can be considered to have a constant period. 
% %We show one such example, HAT-P-3~b, in Figure~\ref{fig:HAT-P-3b}.
% All the fitting plots are shown in Figure~\ref{fig:all_plots_appendix} in the Appendix.

\subsection{Long-term period changes \label{sec:longterm}}
\begin{figure*}[ht!]
\centering
\includegraphics[width=0.95\textwidth]{ 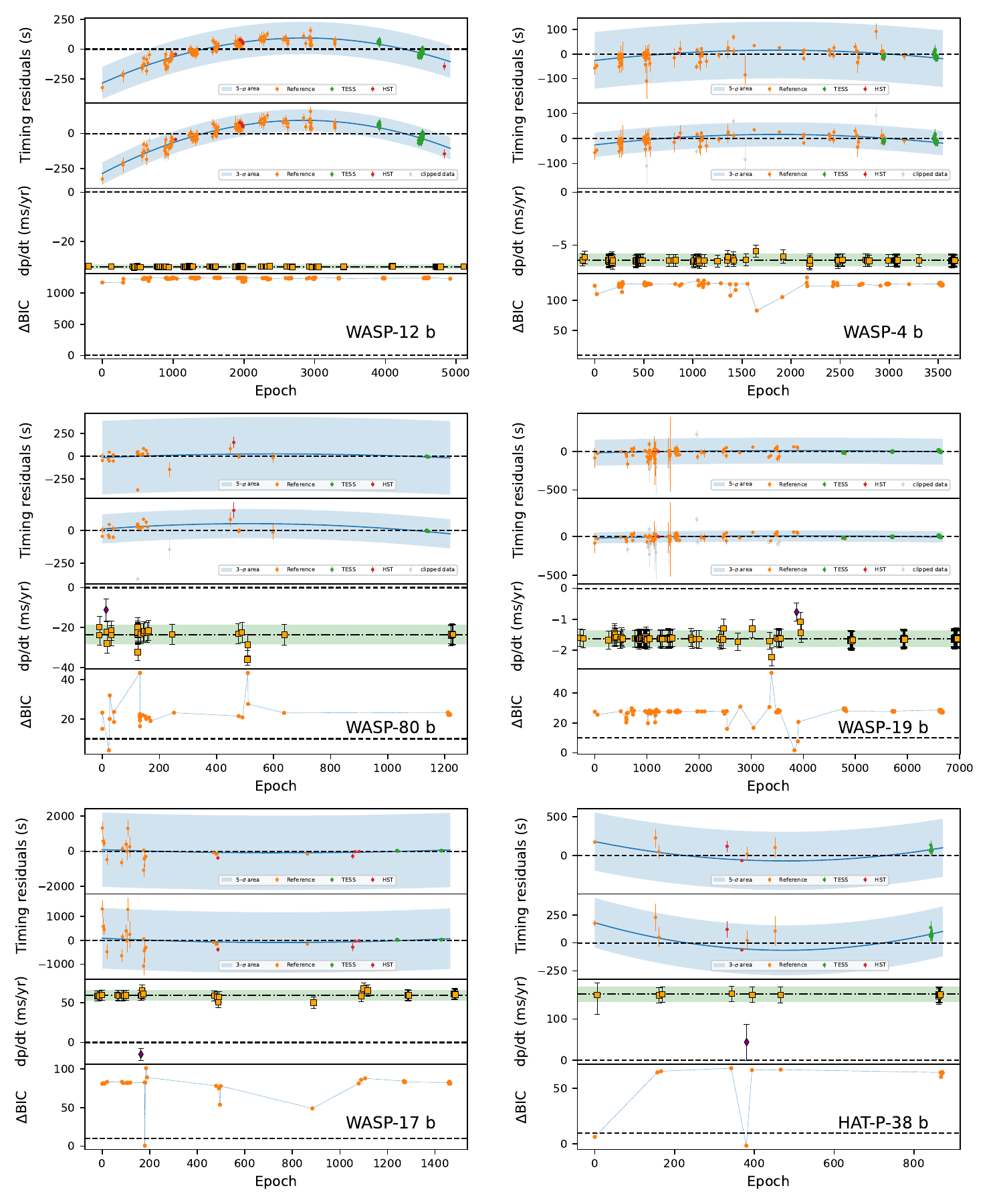}
\caption{
% Time residuals and LOOCV analysis and corresponding $\Delta BIC$ of exoplanets in Section~\ref{sec:longterm}
Time residuals from TTV fitting, LOOCV analysis, and corresponding $\Delta BIC$ of exoplanets studied in Section~\ref{sec:longterm}. For each exoplanet: Top panel: time residual results using 5-$\sigma$ rejection scheme. The second panel: time residual results using 3-$\sigma$ rejection scheme. The blue curves and shaded areas indicate the best-fit quadratic model. Orange points are based on previous literature data. Green points are based on TESS data. Red points are based on HST data from this work. Gray points are clipped data. 
The third panel shows the period change rate $dp/dt$ from fitting a quadratic model after removing each transit timing data point. Orange squares indicate $dp/dt$ values 3-$\sigma$ away from zero, while purple diamonds represent values within 3-$\sigma$. The dash-dotted line marks the original $dp/dt$ fit, with the green shaded area showing the 1$\sigma$ confidence region. The bottom panel displays the corresponding $\Delta BIC$.
}
\label{fig:4.1}
\end{figure*}

% Some special and interesting systems. 
% \subsubsection{Long-term period changes}
We initially have 6 systems that show strong evidence of possible long-term period variation. However, only 2 of them have passed the LOOCV test. All their fitting and LOOCV test results are shown in Figure~\ref{fig:4.1}.
%In this section, we introduce some systems that probably exhibit long-term period changes using archival, TESS, and HST data, with the initial 2 candidates successfully passing the LOOCV test. 
% We refitted these systems with expanding errors for some heavily influential points. 
% Besides, we also refitted the trend using HST and TESS data. The results show most of the trend can not be considered robust. In conclusion, WASP-4b and WASP-12b exhibit a robust long-term decreasing trend.

% Xie: caption 过于简洁。建议以一个系统为例简单的介绍一下上中下4个panel。另外BIC 那个panel 的横虚线是什么？为什么后面几个系统的第一和第二个panels 感觉完全一样 （至少legend 一样）？
% xinyue: 横虚线是all data拟合的best fit pdot，后面几个系统有些3sigma和5sigma两种rejection区分出的点差不多
% The orange points are based on literature data. The green points are based on TESS data. The gray points are clipped data.
% The top panel displays the period change rate dp/dt obtained by fitting the quadratic model after the removal of each single transit timing data. The orange squares show the dp/dt values that satisfy the criterion of being 3σ away from zero, while the purple diamonds represent dp/dt values that fail to meet this criterion. The dash-dotted line marks the original best-fitting dp/dt value before the removal of any data, and the green shaded area marks the corresponding 1σ confidence region. The bottom panel displays the corresponding ∆BIC, where it is evident that the three data points failing the 3σ test also do not satisfy the criterion of ∆BIC< 10.师姐的描述
\begin{itemize}
    \item \textbf{WASP-12~b} is an ultra hot Jupiter with mass of $1.5~M_{\rm Jup}$ and orbital period of 1.1~day around a later-F star \citep{Hebb2009,Collins2017}. 
    Previous TTV studies have indicated a decreasing trend in its orbital period \citep{Maciejewski2016,Patra2017}. 
    \citet{Yee20} presented new transit and occultation observations that provide more decisive evidence for the orbital decay of WASP-12~b. \citet{Turner21} analyzed data from TESS to verify that WASP-12b's orbit is indeed changing with an updated decay rate of $32.53 \pm 1.62$~ms/yr. {\citet{bai2022-study} also found a significant change in the planetary orbit using telescopes of Yunnan Observatories.} \citet{Ivshina22} reported  $\dot{P} = -30.27 \pm  1.11$~ms/yr and \citet{Wang24} found $\dot{P} = -30.19 \pm  0.92$~ms/yr. In this study, by incorporating HST data, we obtain $\dot{P} = -30.31 \pm 0.85$~ms/yr and $\Delta BIC = 1241$, further confirming its orbital decay nature. 
    {We estimate its $Q'_* > 1.7 \times 10^5$ using Equation (1) of \cite{Patra2020}. }
    % and $\dot{P} = -30.92 \pm 7.12$ ms/yr with our HST data individually.
    %  The complete dataset, comprising both HST and TESS observations, as well as HST data individually, all exhibit a consistent trend of period decrease, with the rate of change remaining largely consistent across the different datasets (See Figure~\ref{fig:WASP-12b}). The LOOCV test results appear to be satisfactory, indicating the robustness of our analysis.
    % \begin{figure}[ht!]
    % \plotone{ WASP-12b-e.pdf}
    % \caption{Timing resduals of WASP-12~b. The original HST data with archival data yields $\dot{P} = -30.30 \pm 0.88$ ms/yr. Additionally, $\dot{P} = -30.36 \pm 2.95$ ms/yr with HST data individually while $\dot{P} = -32.83 \pm 1.96$ ms/yr with HST and TESS data.
    % \label{fig:WASP-12b}}
    % \end{figure}

    % \begin{figure}[ht!]
    % \plotone{ WASP-12b-dBIC.pdf}
    % \caption{The LOOCV results of WASP-12~b
    % \label{fig:WASP-12b-dBIC}}
    % \end{figure}
    % xinyue：WASP-4有很多重复epoch，其中重复的HST点已去掉
    % Xie:这里是否排除了其他可能（比如Apsidal precession），而confirm 了 orbital decay？
    % 没有confirm
    \item \textbf{WASP-4~b} is a $1.2~M_{\rm Jup}$ hot Jupiter with a 1.3~day orbital period around a G7V star \citep{Wilson2008}. \citet{Bouma2019} identified an orbital period decay rate of about 10~ms/yr in WASP-4~b, using data from TESS and ground-based transit observations. Based on new radial-velocity  (RV) measurements and speckle imaging observation, \citet{Yee20} inferred that the decrease is most likely caused by the line-of-sight acceleration of the system. \citet{Turner2022} examined TESS, RV, and archival transit data, revealing no Earthward acceleration in the full RV dataset. They suggested a possible additional planet in the system, but its presence could not explain the observed period decay. Instead, they suggested there exists either an orbit decay with $\dot{P} = -7.33 \pm  0.71$~ms/yr or apsidal precession in the system. %\citet{2023Univ....9..506H} have analyzed TESS and archival data of WASP-4b, accounting for the light-time effect  (LTE) from a second companion, using three MCMC models across different timing corrections, but find no conclusive cause for WASP-4b’s apparent transit timing variations. 
    \citet{2023Univ....9..506H} analyzed TESS and archival data of WASP-4~b but found no conclusive cause for its apparent transit timing variations. 
    Using TESS data, \citet{Ivshina22} obtained $\dot{P} = -5.81 \pm  1.58$~ms/yr and \citet{Wang24} obtained $\dot{P} = -6.43 \pm  0.55$~ms/yr for WASP-4~b. Here in this study, by adding HST data to the fitting, we find $\dot{P} = -6.46 \pm 0.58$~ms/yr and $\Delta BIC = 127$, which confirms the orbital decay. {We estimate its $Q'_* > 5.7 \times 10^4$ using Equation (1) of \cite{Patra2020}.}
    % we find $\dot{P} = -6.46 \pm 0.58$ ms/yr using the $5-\sigma$ rejection scheme and $\dot{P} = -6.46 \pm 0.59$ ms/yr using the $3-\sigma$ rejection scheme, which confirms the orbital decay.
    % xinyue:别的地方都没怎么提3sigma,这里要也删掉吗？
    % pass the LOOCV test.
    % \begin{figure}[ht]
    % \plotone{ WASP-4b.pdf}
    % \caption{The fitting results of WASP-4~b
    % \label{fig:WASP-4b}}
    % \end{figure}
    
    % \begin{figure}[ht!]
    % \plotone{ WASP-4b-dBIC.pdf}
    % \caption{The LOOCV results of WASP-4~b
    % \label{fig:WASP-4b-dBIC}}
    % \end{figure}

    \item \textbf{WASP-80~b} is a planet with a mass of $0.54~M_{\rm Jup}$ and an orbital period of 3.1~day around a cool dwarf star \citep{Triaud2013,Triaud2015}. Our analysis shows a decay in the orbital period at a rate of $-23.58 \pm 4.95$~ms/yr, and $\Delta BIC = 19$. LOOCV test results indicate that one key data point significantly affects the TTV fitting result. 
    % Using the earliest data from the discovery paper (considering two versions of the earliest discovery data) and processed data from both HST and TESS, the fit shows a greater decreasing trend while this trend does not meet the 3-sigma criterion ($-115.59 \pm 44.44$ ms/yr). 
    We then refit the model after inflating the error bar of this data point to 3 times its original value, which gave a $\dot{P}$ result of $-10.80 \pm 5.65$~ms/yr, more consistent with a constant period model. 
    \item \textbf{WASP-19~b} is a $1.1~M_{\rm Jup}$ planet with a 0.79~day orbital period orbiting a G8-type star \citep{Hebb2010,Wong2016}. \citet{Patra2020} found strong evidence of orbital decay, but disputed later by studies like \citet{Petrucci20} and \citet{Wang24}. The overall data fitting initially shows a slight period decrease trend, with a rate of $ -1.64 \pm 0.28 $~ms/yr and $\Delta BIC = 33$. 
    % LOOCV testing indicates that two data points influence the result's stability. 
    % When fitting the combined HST and TESS data with the discovery paper data, the discovery paper's data lies outside the 7-sigma range. 
    After re-fitting data with the error bars enlarged, the trend disappears ($ \dot{P} = -0.26 \pm 0.30 $~ms/yr). Thus, we need to continue monitoring this target in the future.
    % no pass
    % \begin{figure}[ht!]
    % \plotone{ WASP-19b.pdf}
    % \caption{The fitting results of WASP-19~b
    % \label{fig:WASP-19b}}
    % \end{figure}
    
    % \begin{figure}[ht!]
    % \plotone{ WASP-19b-dBIC.pdf}
    % \caption{The LOOCV results of WASP-19~b
    % \label{fig:WASP-19b-dBIC}}
    % \end{figure}

    \item  \textbf{WASP-17~b} is an ultra-low-density planet with a mass of $0.49~M_{\rm Jup}$ and a radius of $2.0~R_{\rm Jup}$, which orbits an F6-type star with sub-solar metallicity. It has an orbital period of 3.7~days \citep{Anderson2010,Anderson2011}.
    The period initially appears to be increasing with a best-fit rate of $ 59.01 \pm 6.56 $~ms/yr and $\Delta BIC = 82$. 
    % LOOCV testing shows that the result is heavily influenced by one specific data point. Therefore, we re-fitted the model using only the discovery data and the HST and TESS data separately. Additionally, 
    We then perform the LOOCV test, which identifies a key data point that strongly affects the fitting result. When re-fit the data with this key data point's error bar enlarged by a factor of three, we find no strong evidence supporting this long-term orbital period variation, with $ \dot{P} = 19.63 \pm 7.17 $~ms/yr.
    % to 0.00024~days
    % Result shows constant period.
    % \begin{figure}[ht!]
    % \plotone{ WASP-17b.pdf}
    % \caption{The fitting results of WASP-17~b
    % \label{fig:WASP-17b}}
    % \end{figure}
    
    % \begin{figure}[ht!]
    % \plotone{ WASP-17b-dBIC.pdf}
    % \caption{The LOOCV results of WASP-17~b
    % \label{fig:WASP-17b-dBIC}}
    % \end{figure}

    % 去掉前人处理的HST数据能拟合出递增趋势，但是用前人处理的数据拟合出来是递减的趋势，合起来没趋势
    \item \textbf{HAT-P-38~b} is a planet with a mass of 0.27~$M_{\rm Jup}$ and an orbital period of 4.6~day around a late G star \citep{Sato2012}. 
    It belongs to a rapidly increasing population of transiting Saturn-like planets, and transmission spectra observation has revealed a relatively clear atmosphere with a clear detection of water \citep{Bruno2018}.
    % We find the orbital period is increasing with a rate of $ 161.10 \pm 18.79 $ and $\Delta BIC = 66$. However, we warn that when adding HST data from \citet{Bruno2018}, the trend disappears ($ 11.22 \pm 17.71 $~ms/yr). When re-fit the data with HST from \citet{Bruno2018} and without our data, we find $ -62.78 \pm 18.48 $~ms/yr and $ \Delta BIC = 9$.
    We find that the orbital period is increasing at a rate of $161.10 \pm 18.79$~ms/yr, with a $\Delta BIC$ of 66. 
    However, the LOOCV test shows one of our two HST data points strongly affects the fitting result. When re-fitting after rejecting this HST data point, we find a result consistent with a constant period model, with $\dot{P} = 42.63 \pm 45.93$. Thus, Further observations are needed to verify this trend.
    %However, we note that when incorporating HST data from \citet{Bruno2018}, this trend disappears ($\dot{P} = 11.22 \pm 17.71$~ms/yr). 
    %Re-fitting the data using only HST data from \citet{Bruno2018} and excluding our own data, we find a rate of $-62.78 \pm 18.48$~ms/yr, with $\Delta BIC = 9$.
    % xinyue: 上面最后一句要提吗
    % In our fitting process, removing the previously processed HST data allows us to fit an increasing trend; however, when using the earlier processed data, we obtain a decreasing trend. When combined, there is no clear trend.
    %xinyue: 这段话需要你补充和重写下，逻辑要通顺些。
    %xinyue: 老师这个源放的是只用我们HST数据的图，然后inspiral time那里用的是全部数据拟合的pdot，这是可以的嘛

    % In our analysis, we incorporated the previously processed HST data from the same raw data\citep{Bruno2018}applying an error expansion to match the TESS error bars. The fitting results remain consistent with previous analyses, indicating a constant period.
    
    % \begin{figure}[ht!]
    % \plotone{ HAT-P-38b.pdf}
    % \caption{The fitting results of HAT-P-38~b
    % \label{fig:HAT-P-38b}}
    % \end{figure}
    % \begin{figure}[ht!]
    % \plotone{ HAT-P-38b-o.png}
    % \caption{The fitting results of HAT-P-38~b using previous HST data
    % \label{fig:HAT-P-38b-o}}
    % \end{figure}

\end{itemize}

\subsection{Short-term TTV \label{sec:shortterm}}
\begin{figure*}[ht!]
\centering
\includegraphics[width=0.95\textwidth]{ 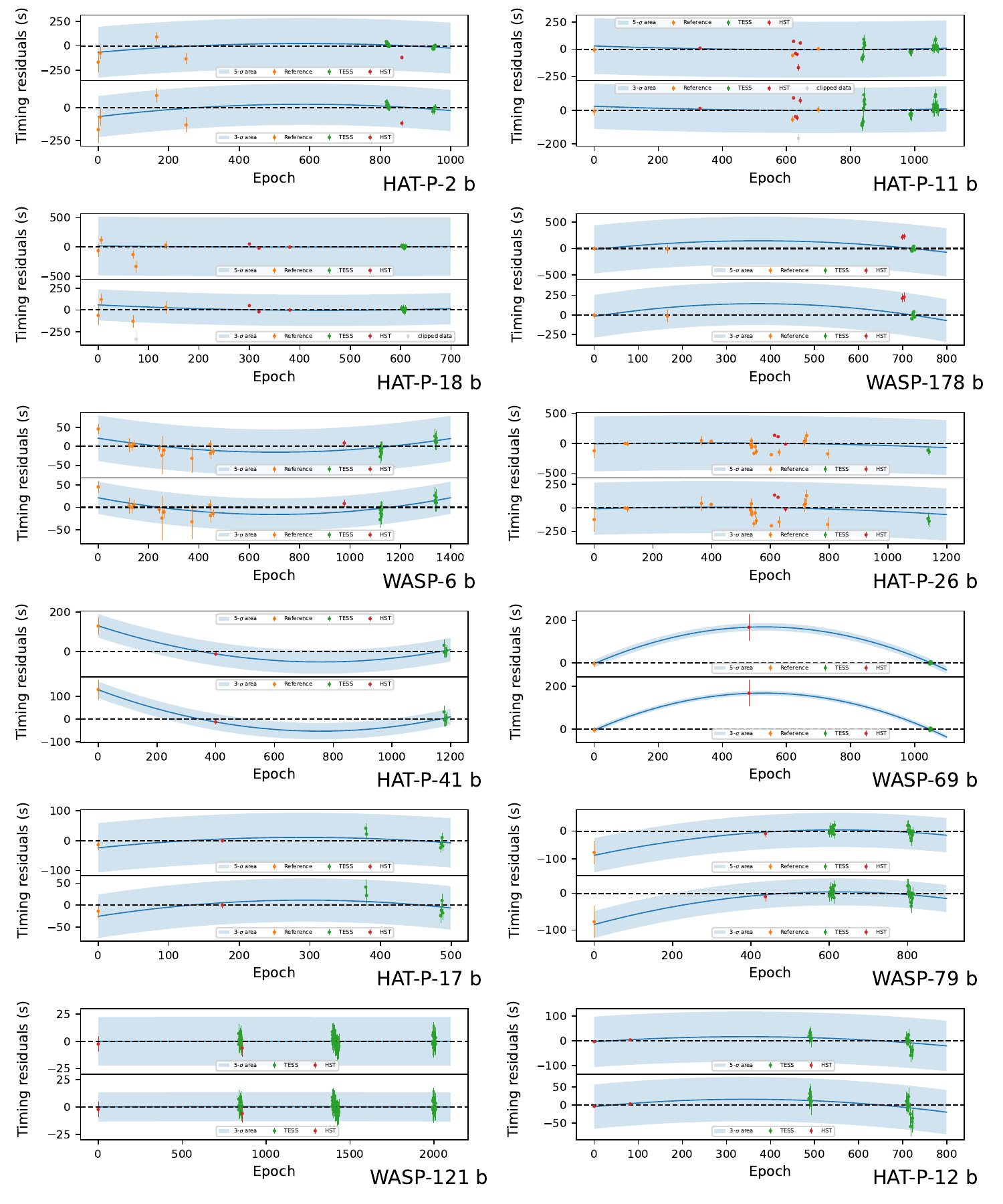}
\caption{
Time residuals of exoplanets in Section~\ref{sec:shortterm} and Section~\ref{sec:interest}
\label{fig:4.2}
}
\end{figure*}

%We have four planetary systems showing possible short period variations. 
% Although some systems do not show trend in the long-term orbital period variation, they may show possible short-term transit timing variations. We here present four such systems in this section.
% These systems worth further studies to find our the exact reasons behind these variations.
Although some systems do not exhibit trends in long-term orbital period variations, they may show potential short-term transit timing variations  (TTVs). In this section, we present four such systems. All their fitting results are shown in Figure~\ref{fig:4.2}. These systems are worth further study to identify the exact reasons behind these variations. 

%list some planets that indicate short-term significant variability, although they show constant periods in the long-term TTVs fitting. 
% They worth more study to find the exact reason for these phenomena.

\begin{itemize}
    \item \textbf{HAT-P-2~b}  (also called HD~147506~b) is a hot Jupiter wit mass of $9.04~M_{\rm Jup}$, eccentricity of 0.52, and orbital period of 5.6~day, which orbits around a bright F8 star \citep{Bakos2007}. 
    The study of \citet{Jacobs24} reveals the rapid heating and cooling of this highly eccentric hot Jupiter, which they attribute to the possible star-planet interaction. \citet{deBeurs23} have studied its orbital period evolution using simulation and suggested further monitoring it with precise RVs and transit and eclipse timings. 
    Our fitting results suggest the potential presence of short-term TTVs with an amplitude as large as 100~s, which may be related to the system's strange large eccentricity.
    % \begin{figure}[ht!]
    % \plotone{ HAT-P-2b.pdf}
    % \caption{The fitting results of HAT-P-2~b
    % \label{fig:HAT-P-2b}}
    % \end{figure}
    
    \item \textbf{HAT-P-11~b} is a $0.081~M_{\rm Jup}$ planet with an orbital period of 4.9~day, orbiting a bright and metal rich K4-type dwarf star \citep{Bakos2010}. \citet{Yee2018} discovered a second planet HAT-P-11~c in this system. With this additional companion, the eccentricity and stellar obliquity of HAT-P-11~b could be explained under the assumption that HAT-P-11~c was also misaligned. \citet{Lu2024} proposed a two-step dynamical process that can reproduce all observed properties of this system. Our TTV fitting results show that short-term strong variations exist in this system, which awaits future confirmation.
    
    % \begin{figure}[ht!]
    % \plotone{ HAT-P-11b.pdf}
    % \caption{The fitting results of HAT-P-11~b
    % \label{fig:HAT-P-11b}}
    % \end{figure}
    
    \item \textbf{HAT-P-18~b} is a $0.20~M_{\rm Jup}$ planet with an orbital period of 5.5~day, orbiting a K2 dwarf star \citep{Hartman2011_18_19}. Our fitting results indicate that neither the long-term constant-period model nor the quadratic model provides a satisfactory fit, suggesting the possible presence of short-term TTVs with an amplitude as large as 100~s in this system.
    % \begin{figure}[ht!]
    % \plotone{ HAT-P-18b.pdf}
    % \caption{The fitting results of HAT-P-18~b
    % \label{fig:HAT-P-18b}}
    % \end{figure}
    
    \item \textbf{WASP-178~b} is a bloated planet with a mass of $1.7~M_{\rm Jup}$, radius of $1.81~R_{\rm Jup}$, and orbital period of 3.3~day, which orbits around an A1V-type star with an effective temperature of $9350$~K \citep{Hellier2019}. 
    Our fitting results indicate that neither the long-term constant-period model nor the quadratic model provides a satisfactory fit, suggesting the possible presence of short-term TTVs with an amplitude as large as 250~s in this system.
    %Our fitting results show that both the long-term constant period model and quadratic model could not yield a good fit, demonstrating the potential presence of short-term TTV in this system.
    % \begin{figure}[ht!]
    % \plotone{ WASP-178b.pdf}
    % \caption{The fitting results of WASP-178~b
    % \label{fig:WASP-178b}}
    % \end{figure}
\end{itemize}

\subsection{Special Interest \label{sec:interest}}
In this section, we present 8 systems that are classified as `interesting' by us, including planets showing possible weak period variation, and targets showing significantly different results between this study and the literature. All their fitting results are shown in Figure~\ref{fig:4.2}. 
All of these targets warrant further observations.
 %Planets have the potential period change, which means they do not meet both two criteria. 
%Some of them show constant periods using all data but indicate a decreasing trend when using both HST and TESS data. 
%Besides, we also noticed some transit times of planets that have too large change to beyond 5 sigma fitting. 

\begin{itemize}
    \item \textbf{WASP-6~b} is an inflated sub-Jupiter mass planet, with with a mass of $0.50~M_{\rm Jup}$ and radius of $1.2~R_{\rm Jup}$,   indicating that it has a puffed-up, extended atmosphere. It orbits around a mildly metal-poor solar-type star with a 3.4~day orbital period \citep{gillon_discovery_2009}. The orbital period was found to be increasing by \citet{Wang24}, and adding new HST data further constrains the rate of orbital period change. Our analysis yields a rate of $ 16.21 \pm 4.77 $~ms/yr and $\Delta BIC = 9$, which is close to the value obtained by \citet{Wang24} ($ \dot{P} = 20.66 \pm 5.19 $~ms/yr).
    % The period change shows an overall increase of $ 16.21 \pm 4.77 $ ms/yr. 
    %Adding new HST data reduces this rate comparable to the result of \citet{Wang24}. 
    The LOOCV test indicates that the first data point has greatly affected the fitting result. We then refit the model after inflating the error bar of this data point to 3 times its original value, which shows a $\dot{P}$ value of $11.99 \pm 5.27$~ms/yr, consistent with a constant period model. 
    % LOOCV reveals a strong dependence on the discovery paper's data. When its errors are tripled, the period change rate drops to \ ( 15.00 \pm 5.76 \) ms/yr and no longer meets the 3-sigma significance.
    % no pass
    % \begin{figure}[ht!]
    % \plotone{ WASP-6b.pdf}
    % \caption{The fitting results of WASP-6~b
    % \label{fig:WASP-6b}}
    % \end{figure}
    
    % \begin{figure}[ht!]
    % \plotone{ WASP-6b-dBIC.pdf}
    % \caption{The LOOCV results of WASP-6~b
    % \label{fig:WASP-6b-dBIC}}
    % \end{figure}
    % 去掉之前人处理的HST点之后拟合不出符合判定的趋势
    \item \textbf{HAT-P-26~b} is a low-density planet with a mass of $0.059~M_{\rm Jup}$ and an orbital period of 4.2~day around a K1-type dwarf star \citep{Hartman2011_26}. Its mass, comparable to Neptune and Uranus, coupled with a radius approximately 65\% larger than those of Neptune and Uranus, sets HAT-P-26b apart from other transiting Super-Neptunes \citep{Hartman2011_26}. 
    HAT-P-26b exhibits a well-constrained heavy element abundance, which is lower than that observed in Uranus and Neptune \citep{Wakeford2017,MacDonald2019}. {\citet{a-thano2023-revisiting} suggested the $1.98 \pm 0.05$ minute amplitude signal in the TTV analysis could be due to a $0.02~M_{\rm Jup}$ planet in a 1:2 resonance orbit.}
    When we use our processed HST data along with literature data, fitting shows a constant period  ($\dot{P} = -20.60 \pm 11.72$~ms/yr). However, if we fit the data with previous HST data, we observe a significant increasing trend in the period ($\dot{P} = 119.64 \pm 9.12$~ms/yr).
    % 改成了最新一次拟合的结果
    % Fitting all data shows an exaggerated periodic increase  (\ ( 119.84 \pm 9.15 \) ms/yr, $\Delta$BIC = 180) heavily influenced by only 1 or 2 points; expanding HST errors and combining with TESS and discovery data reveals a 3-sigma consistent decreasing period  ($\Delta$BIC = 9.7); tripling discovery data errors results in a constant period  (\ ( -120.02 \pm 91.06 \) ms/yr), illustrating the significant impact of optimistic error estimates on fitting outcomes.
    %     no pass
    % There is a trend of period increase, but some of the data significantly affect this trend and require further analysis.
    % \begin{figure}[ht!]
    % \plotone{ HAT-P-26b.pdf}
    % \caption{The fitting results of HAT-P-26~b
    % \label{fig:HAT-P-26b}}
    % \end{figure}
    
    % \begin{figure}[ht!]
    % \plotone{ HAT-P-26b-dBIC.pdf}
    % \caption{The LOOCV results of HAT-P-26~b
    % \label{fig:HAT-P-26b-dBIC}}
    % \end{figure}
    
    % boma: may be moving to interesting candidate
    % xinyue：之前用的是wakefordcombined之后的数据，同时也加上分别的数据就会显示常数周期，combine的时候还包括2012Hartman的点和我处理的hst，那到底应该用哪些数据呢？暂时按以前的图了。
    % 这个源现在放的图是只有一个文献点的，inspiral time用的是加上了更多combine之前的点拟合的pdot。
    \item \textbf{HAT-P-41~b} is a $0.80~M_{\rm Jup}$ planet with an orbital period of 2.7~day, orbiting a moderately bright (V=12.4) F star \citep{Hartman2012}. 
    % For this target, we have also incorporated transit time measured by \citet{Wakeford2020} from the HST/UVIS G280 Grism instrument.
    We initially identify an increasing trend in its orbital period using archival data and our TESS and HST data, with $\dot{P} = 86.32 \pm 27.08$~ms/yr and $\Delta BIC = 8$. We later find the only one archival data used in the fitting from \citet{Wakeford2020} is actually a combined epoch data from many observations of Spitzer and HST spanning 10~years. 
    We then go ahead to use individual timing data from literature instead of this one combined data \citep{Hartman2012, Wakeford2020}, and re-fit the TTV model. The increasing trend disappears at this time, with $\dot{P}$ as $-7.68 \pm 13.07$~ms/yr.
    %\textcolor{red}{For this target, we also incorporated the transit times measured by \citet{Wakeford2020} using the HST/UVIS G280 Grism and HST/STIS G430L/G750L Grism instruments. Initially, we used the combined data point from \citet{Wakeford2020}, which includes measurements from \citet{Hartman2012}, HST, and Spitzer.}
    %The fitting suggests the orbital period has an increasing trend with $\dot{P} = 86.32 \pm 27.08$~ms/yr, but $\Delta BIC = 8<10$. 
    %\textcolor{red}{When using the original data points instead of the combined data point, the fit yields a $\dot{P}$ value of $-7.68 \pm 13.07$~ms/yr, consistent with a constant period model.}

    % when the HST data errors are increased, the period trend only meets the 2-sigma range; further, expanding the discovery data error to three times its original value, the period trend fits within only the 1-sigma range.
    % no pass
    % \begin{figure}[ht!]
    % \plotone{ HAT-P-41b.pdf}
    % \caption{The fitting results of HAT-P-41~b
    % \label{fig:HAT-P-41b}}
    % \end{figure}

    % 只满足2sigma，点太少了
    \item \textbf{WASP-69~b} is a bloated Saturn-mass planet with a mass of $0.26~M_{\rm Jup}$, a radius of $1.1~R_{\rm Jup}$, and an orbital period of 3.9~day  around an active, mid-K-type dwarf star \citep{Anderson2014}. 
    %boma:貌似有人研究它的大气逃逸？大气损失逃逸是不是可以增加pdot (可以和陈章亮讨论下）。
    Recently, through X-ray and extreme ultraviolet  (XUV) bands research, \citet{Levine2024} discovered that WASP-69~b may experience atmospheric dissipation. In our study, the decreasing period trend only meets a 2 sigma criterion with $\dot{P} = -117.92 \pm 41.77$~ms/yr and does not satisfy the $\Delta BIC$ criterion, suggesting a weak trend awaiting further confirmation. 
    % $\Delta BIC = 5$
    % Because the trend strongly correlates with the added HST data, we included the same HST data processed by Tsiaras. After adding times from Tsiaras  (2457617.14615 ± 0.00024), the data still indicate a decreasing period and meet both the deltaBIC and 3 sigma criteria ($\dot{P} = -64.36 \pm 15.09 $ ms/yr). 
    % After enlarging the errors in the discovery data, the decreasing trend is still well-supported ($\dot{P} = -64.60 \pm 14.87 $ ms/yr).
    % \begin{figure}[ht!]
    % \plotone{ WASP-69b.pdf}
    % \caption{The fitting results of WASP-69~b
    % \label{fig:WASP-69b}}
    % \end{figure}

    % 点比较少，不满足判定，可考虑删掉，不提
    \item \textbf{HAT-P-17~b} is a hot Jupiter with a mass of $0.53~M_{\rm Jup}$ and an orbital period of 10~day  around an early K-type dwarf star \citep{Howard2012}. 
    % It is situated in a multi-planet system, with the inner planet exhibiting an eccentric, short-period orbit, while the outer planet HAT-P-17~c is a cold Jupiter with a nearly circular orbit and a period of 4.4 years \citep{Howard2012}. 
    It is situated in a multi-planet system. The inner planet has an eccentric, short-period orbit, while the outer planet, HAT-P-17~c, is a cold Jupiter with a nearly circular orbit and a period of 4.4~years \citep{Howard2012}.
    While our analysis indicates a potential decreasing trend with $\dot{P} = -29.43 \pm 15.72$~ms/yr, the limited data available prevents the results from meeting the statistical 'strong' criteria.
    % \begin{figure}[ht]
    % \plotone{ HAT-P-17b.pdf}
    % \caption{The fitting results of HAT-P-17~b
    % \label{fig:HAT-P-17b}}
    % \end{figure}

    %满足3sigma但其实deltaBIC甚至小于0, delete发现文章数据，加入Brown et al.  (2017)数据后不满足判定
    \item \textbf{WASP-79~b} is a highly-bloated planet with a mass of $0.9~M_{\rm Jup}$ and orbital period of 3.7~day around an F-type star \citep{Smalley_WASP7879_2012}.
    % There is a decreasing trend, but it does not meet the $\Delta BIC$ criterion and is mainly strongly correlated with the data from the discovery paper. 
    Our analysis initially shows a decay in the orbital period at a rate of $-78.80 \pm 26.22$~ms/yr, and $\Delta BIC = -1$. 
    We then re-fit the model after removing the first data point and adding data from \citet{Brown2017}. This yields a $\dot{P}$ value of $-47.81 \pm 19.76$~ms/yr, which is consistent with a constant period model.
    % After amplifying 3 times the error of discovery paper data, it indicated a constant period. 
    % \begin{figure}[ht!]
    % \plotone{ WASP-79b.pdf}
    % \caption{The fitting results of WASP-79~b
    % \label{fig:WASP-79b}}
    % \end{figure}

    % 只有7sigma因为包含了discover点所以能满足deltaBIC，但这个源本身可以作为说明高精度数据重要性的例子
    % 之前的数据离散程度太大都在5sigma之外，只用tess,hst就可以拟合常数周期
    \item \textbf{WASP-121~b} is a planet with a mass of $1.2~M_{\rm Jup}$ and an orbital period of a 1.3~day around an active F-type star \citep{Delrez2016}.
    The fitting results show the orbital period is increasing. The fits using both the 3-sigma and 5-sigma rejection schemes do not meet the $\Delta BIC < 10$ criterion, but the fitting using a 7-sigma rejection scheme yields a result that meets the $\Delta BIC > 10$ criteria. 
    However, if using only data from HST and TESS which we deemed trustworthy, we found a $\dot{P}$ of $0.02 \pm 2.01$~ms/yr, showing no evidence of significant period variation. This result also demonstrates the superiority of two HST data points compared to the tens of data points from ground-based telescopes. 
    %From the figure, we can find that the HST data we incorporated played an effective role in filtering out data with a dispersion during the fitting process.
    % When we refitted the data with the enlarged error of discovery data and HST data, this trend disappeared.
    % \begin{figure}[ht!]
    % \plotone{ WASP-121b.pdf}
    % \caption{The fitting results of WASP-121~b
    % \label{fig:WASP-121b}}
    % \end{figure}
    % \begin{figure}[ht!]
    % \plotone{ WASP-121b-dBIC.pdf}
    % \caption{The LOOCV results of WASP-121~b
    % \label{fig:WASP-121b-dBIC}}
    % \end{figure}

    % 都不满足deltaBIC
    % \item \textbf{WASP-127 b} is a $0.18~M_{\rm Jup}$ heavily inflated planet with a 4.2-day orbital period around a bright (V=10.16) G5-star\citep{Lam2017}.
    % The 5-sigma fit shows a constant period, while the 3-sigma fit indicates a decreasing period. After enlarging the error bars of the first and second points, the trend disappears.
    % \begin{figure}[ht!]
    % \plotone{ WASP-127b.pdf}
    % \caption{The fitting results of WASP-127~b
    % \label{fig:WASP-127b}}
    % \end{figure}
    
    % \begin{figure}[ht!]
    % \plotone{ WASP-127b-dBIC.pdf}
    % \caption{The LOOCV results of WASP-127~b
    % \label{fig:WASP-127b-dBIC}}
    % \end{figure}

    % 全部数据没趋势，只用HST和TESS的趋势也不满足判定
    \item \textbf{HAT-P-12~b} is a low density planet with a mass of $0.21~M_{\rm Jup}$ and orbital period of 3.2~day around a K4-type dwarf star \citep{Hartman2009}. The fitting with all available timing data indicates it has a constant orbital period. 
    % The results also show a constant period when fitting the discovery paper data together with HST and TESS data. 
    However, when fitting only the HST and TESS data, we find the orbital period is decreasing with $ \dot{P} = -40.22 \pm 17.92$~ms/yr.   
    % In summary, our analysis supports the hypothesis that the period is constant. 
    This new fitting does not yet satisfy the 3-$\sigma$ and $\Delta BIC > 10$ criterion. 
    Future high-precision transit timing data are needed to confirm this decreasing trend. % of this system.
    
    % \begin{figure}[ht!]
    % \plotone{ HAT-P-12b-0.pdf}
    % \caption{The fitting results of HAT-P-12~b
    % \label{fig:HAT-P-12b}}
    % \end{figure}

%boma comment off
    % \textbf{KELT-7 b} is a $1.3~M_{\rm Jup}$ planet with a 2.7-day orbital period around a bright (V=8.54) rapidly rotating ($73\pm 0.5 km/s$) F-star \citep{Bieryla2015}.\textcolor{red}{TBC}
    % % % 之前数据处理部分有问题导致HST数据偏移过大，使用 (Pluriel et al., 2020)处理的HST数据拟合也还是会有很大离散，考虑原因是TESS数据很精确很集中
    % When simulated, the HST data show an anomalous shift  (6 minutes), which is beyond 5 sigma. At 3 sigma, several other data points are also excluded.
    % \begin{figure}[ht!]
    % \plotone{ KELT-7b.pdf}
    % \caption{The fitting results of KELT-7~b
    % \label{fig:KELT-7b}}
    % \end{figure}
    % 常数周期
    % \item \textbf{WASP-96 b} is $0.48~M_{\rm Jup}$ planet with a 3.4-day orbital period around a G8-star\citep{Hellier2014}.
    % There is a decreasing period trend, but it is not satisfie the 3-sigma criterion and does not meet the deltaBIC criterion. After refitting the data from the discovery paper, TESS, and HST, it showed a constant period.
    % % \begin{figure}[ht!]
    % % \plotone{ WASP-96b.pdf}
    % % \caption{The fitting results of WASP-96~b
    % % \label{fig:WASP-96b}}
    % % \end{figure}
\end{itemize}

\section{Discussion}
\subsection{Inspiral Timescale of HJs}
% How long will the planet disappear? 
% $\frac{P}{\dot{P}\pm 3\sigma }$ 

\begin{figure*}[ht!]
\plotone{ 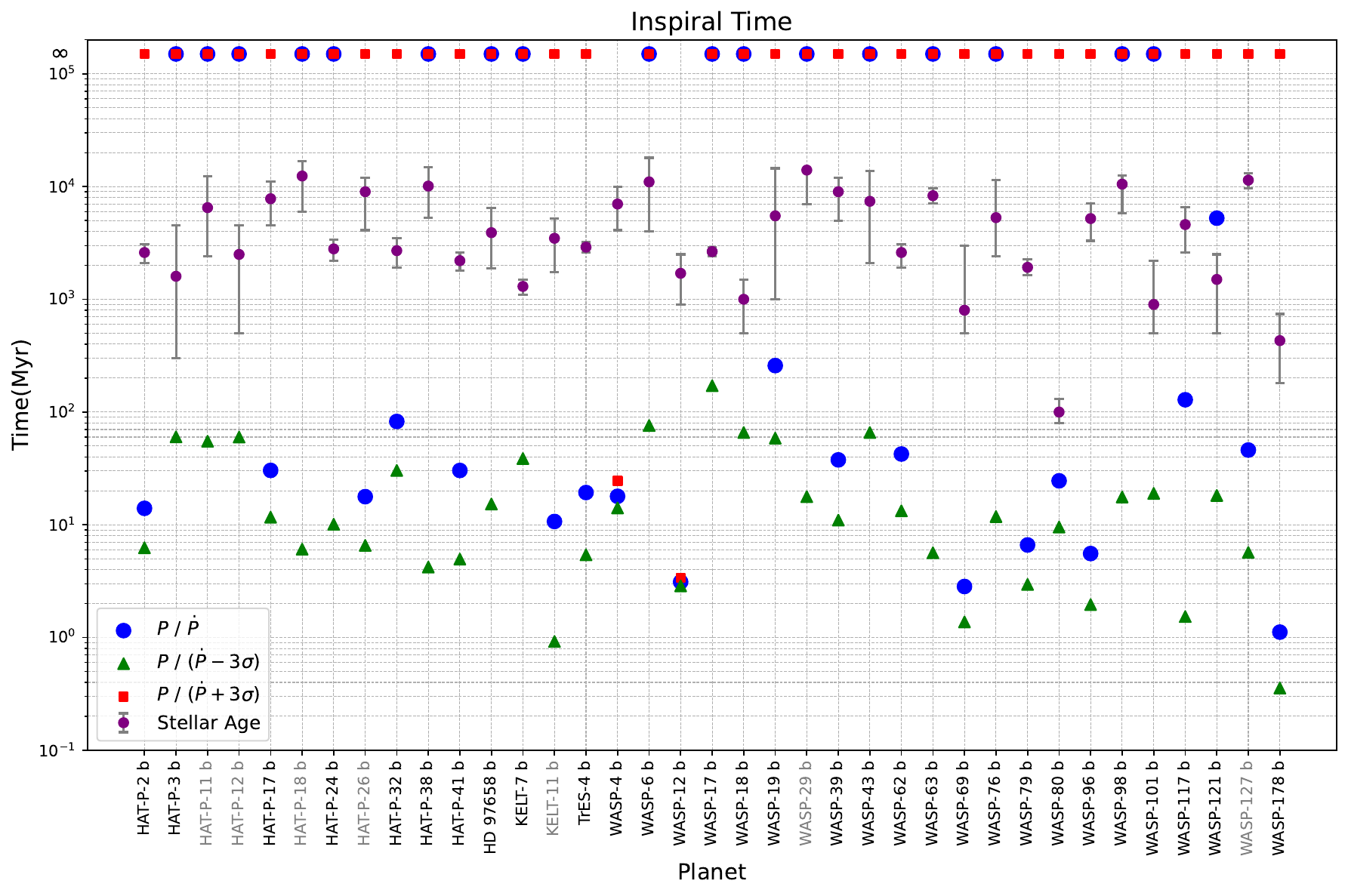}
\caption{Inspiral Timescale for exoplanets in our sample. Grey: exoplanets with mass $M \lesssim 0.25 M_J$; Black: HJ  ($0.25 M_J < M \lesssim 13.6 M_J $). The stellar ages are taken from: \citet{Pal2010,Chan2011,Bakos2010,Hartman2009,Howard2012,Hartman2011_18_19,Kipping2010,Hartman2011_26,Hartman2011_32_33,Sato2012,Hartman2012,2021AJ....162..118E,Bieryla2015,2018ApJ...860..109G,Bonomo2017,2013ApJ...779L..23P,gillon_discovery_2009,Anderson2011,Hellier2009,Hebb2010,Hellier2010,Faedi_wasp39_2011,2022AA...668A..17S,Hellier2012,Anderson2014,West_Three_2016,Smalley_WASP7879_2012,Triaud2013,2023ApJ...944L..56M,Hellier2014,Delrez2016,Lam2017,Rodriguez2020}.
\label{fig:inspiral_time}}
\end{figure*}

The formation and evolution of HJs remain a hot topic in the exoplanet research field \citep{Dawson2018, Zink23, Wu23, Fang23}. 
Here, we can contribute to this topic by putting constraints on the in-spiral time scale of the HJs. Assuming the rate of orbital period decay remains constant, the inspiral timescale  (or known as orbital decay timescale) is usually defined as:
\begin{equation}
T_{\rm inspiral} = \frac{P}{\dot{P}},
\end{equation}
which represents how long it would take for a hot Jupiter to be engulfed by its host star.

%xinyue:图已于11.2更新
%xinyue:HAT-P-38是同时采用了两组HST数据的结果，HAT-P-41是加入了文章里未combine的更多数据的结果
%boma
%1-sigma, 12 *2/3 = 8 dead, *10 =100 dead, 从新生到几十亿年，~1/3 dead
%3-sigma, 3 dead *100 ~300 dead 几十亿年， 减少的速率也蛮大的。
%xinyue:1sigma中包含挺多不能满足deltaBIC>10条件的源的，老师这个*2/3是因为1sigma覆盖68%么

% \begin{figure*}[ht]
% \plotone{ time_inspiral_1sigma.pdf}
% \caption{Time Inspiral ($1\sigma$). The stellar ages are from the same papers as Figure~\ref{fig:inspiral_time}.
% \label{fig:inspiral_time_1}}
% \end{figure*}

% \begin{figure*}[ht]
% \plotone{ time_inspiral_3sigma.pdf}
% \caption{Time Inspiral ($3\sigma$). The stellar ages are from the same papers as Figure~\ref{fig:inspiral_time}.
% \label{fig:inspiral_time_3}}
% \end{figure*}
%目前在画5sigma rejection 的图，优先选全部数据未放大误差的拟合结果，部分存在影响极大 (符合了Pdot判定条件）不合理小误差点使用修正后的结果，放大到原本的3倍5倍 or 1min。
%对于部分行星，数据库第一个点不一定来自于发现文章数据，例如WASP127第一个点来自2020年对2014年的EulerCam数据的处理，而发现文章当时虽然对多个观测项目进行了对比，但只公布了WASP得到的T0，之后要对前文的相关描述进行修正。
In Figure~\ref{fig:inspiral_time}, we present the distribution of $T_{\rm inspiral}$, along with its 3-$\sigma$ upper and lower limits calculated using a 3-$\sigma$ error in $\dot{P}$. The distribution indicates that most host stars are older than 0.5~Gyr, while some HJs exhibit relatively short inspiral timescales. This finding suggests that certain HJs may undergo the engulfment process within the next several Myr. 
To date, nearly 400 hot Jupiters  (HJs) have been discovered. 
Our inspiral timescale analysis further implies that, if the current rate of period change remains constant, approximately $0.5\%$ ($\sim 2/400$) of these HJs will be lost within the next $10^7$~years. With a linear extrapolation, this trend would predict the disappearance of all HJs within the next $2 \times 10^9$~years. Consequently, this analysis suggests that the majority of HJs may be expected to be lost over a timescale comparable to the typical ages of their host stars, which are on the order of several Gyr.

Given the older ages of HJ host stars, 
% a mechanism for continuous HJ formation
% Dichang: 改为：some mechanisms for continuously HJ formation on the timescale up to several Gyr也许更好
some mechanisms for continuous HJ formation on the timescale up to several Gyr
is needed so as to sustain the observed HJ population. This supports the need for a `late-arrived' hot Jupiter population, as proposed by \citet{Hamer2022}, \citet{Chen2023}, and Chen et al. (2024, submitted). 
%这里也可以加上Hamer et al. 2022和Chen et al. 2024 PAST VI.
% 这里的引用应该是 Hamers & Schlaufman 2022; Chen et al. 2024) Hamers & Schlaufman (2022) :http://arxiv.org/abs/2205.00040 Chen et al. 2024 即将投稿。
%Hamer已加，Chen2024要自己编一个bib吗
One possible channel for HJ replenishment is the dynamical high-eccentricity migration \citep{Rasio96, Chatterjee08, Naoz12, Petrovich15}, a scenario supported by recent observations \citep{Zink23}. 
However, observations of known cool and warm Jupiters have shown that very few of them can become HJs through tidal decay only, as this requires a specific combination of eccentricity and semi-major axis. 
%This occurs because warm Jupiters need a particular combination of eccentricity and semimajor axis to approach their host star closely. \textcolor{red}{any references for this?} 
%\citet{Chen2023} found that hot Jupiter host stars are younger than warm/cold Jupiter hosts by $\sim$ 2 Gyr. 
%Besides, they found that the frequency of hot Jupiters declines with age while the frequency of warm/cold Jupiters does not. 
% Our study implies that the number of HJs will decrease, which contradicts the 0.5\% HJs observed today. 
% This means that there may be some other mechanisms that can drive the continuous formation of HJs in a time scale of 100~Myrs, which is different from the dynamical high-eccentricity migration scenario \citep{Rasio96, Chatterjee08}. 
% eccentricity excitation in the postdisk phase
% trigger efficient tidal circularization
It is likely that secular chaotic processes after disk dissipation has pumped the eccentricities of warm/cool Jupiters, setting up stages for high-eccentricity damping, which ultimately leads to HJ formation \citep{Wu11,Hamers17,Teyssandier19}.
This picture is consistent with the scenario proposed by \citet{Wu23}, where they argue the HJ systems are the natural end stage of giant planets with a wide range of eccentricities, originating from interactions either during the disk phase or in the post-disk phase. 
%where they argue the HJ system are natural results of produced from the more violent side of a continuous post-disk dynamical interactions.
%Recently, \citet{wang24} have proposed that a random scatter-scattering process may facilitate the HJ formation. 

%boma: 这里可以读一下，汪师姐昨天讲的，关于温木星的论文， 参考Sgro et al.  (2024, arxiv)。

\subsection{Impact of Error Bar Reporting}

\begin{figure}[ht!]
\plotone{ 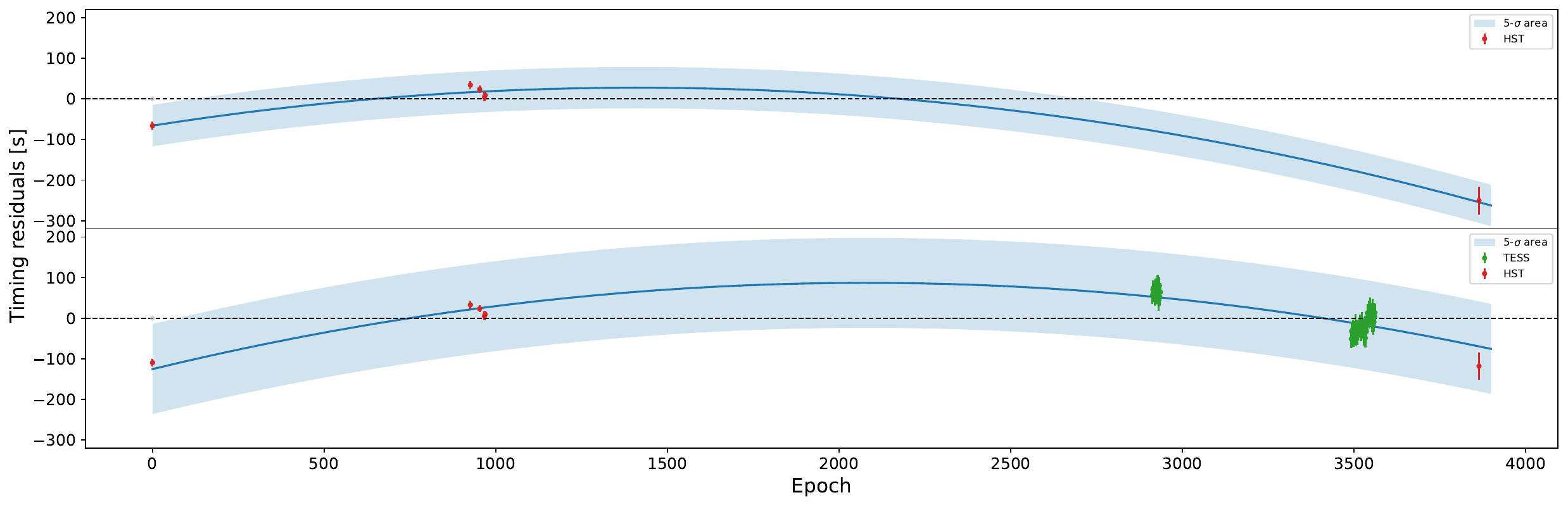}
\caption{Timing residuals when conducting the TTV modeling of WASP-12~b. 
Top panel: Quadratic fitting of the HST data alone.  Bottom panel: Quadratic fitting of the HST and TESS data together. We obtain $\dot{P} = -31.36 \pm 2.95$~ms/yr when fitting the HST data alone, and $\dot{P} = -32.84 \pm 1.78$~ms/yr when fitting the HST and TESS data together. For comparison, the original fitting in Section~\ref{sec:analysis} with all selected archival data yields $\dot{P} = -30.31 \pm 0.85$~ms/yr.
% The original HST data with archival data yields $\dot{P} = -30.31 \pm 0.85$ ms/yr. Additionally, $\dot{P} = -30.36 \pm 2.95$ ms/yr with HST data individually while $\dot{P} = -32.83 \pm 1.96$ ms/yr with HST and TESS data.
\label{fig:WASP-12b}}
\end{figure}
%xinyue： 图8考虑做个新版本，把那个panel 1，panel 3 去掉吧。留下TESS和HST数据的两个panel参考下。
% ok，不确定要不要提，HST和TESS拟合的$\Delta BIC = 330$, HST only $\Delta BIC = 102$，都是能很好地满足deltaBIC判定
    
% \textcolor{red}{In our study, we often obtain significantly smaller error bars for the timing data obtained from HST compared to those from TESS. 
% To investigate the potential presence of additional systematic uncertainties, we artificially inflate the HST error bars to match or exceed those of the TESS observations. This approach is based on the observation that TESS provides comprehensive coverage of the entire transit window, whereas HST typically covers only 50\% to 60\% of the transit event. By adopting this conservative strategy, we aim to ensure that our results are robust and not strongly affected by the limited transit window coverage of HST observation.} 
%boma: 我感觉如果是最开始就把hst的误差棒处理过了，那么需要在最开始的章节就提。
%xinyue：老师我们后来讨论说不直接扩大HST误差了，这段应该不需要了

% Moreover, we have frequently encountered literature timing data reported with unrealistically small error bars, which may not adequately account for all sources of uncertainty. 
% To address this issue, we recommend that future studies adopt a more cautious approach when reporting error bars for transit timing data, and publish the Monte Carlo chain data in their light curve fitting together with the light curve data. 
% Specifically, researchers should consider the potential impact of incomplete transit coverage and other observational limitations, ensuring that the reported uncertainties reflect a realistic assessment of the observational uncertainties. 

In our study, we often encounter literature timing data reported with unrealistically small error bars, which may not adequately account for all sources of uncertainties from observations, especially from the ground-based telescopes. 
To demonstrate the significance of using only data with reliable error bars (those with well-known systematics) in the TTV study, we present one such study using WASP-12~b as an example (also see WASP-121~b). We fit a quadratic model using space-based timing data from HST and TESS observations, and show the results in Figure~\ref{fig:WASP-12b}. We consider these data as data with the most trustworthy error bars reported. 
Using only six data points from HST, we obtain a fitting result of $\dot{P}$ as $-31.36 \pm 2.95$~ms/yr, compared to $-30.31 \pm 0.85$~ms/yr when using all 190 data points from literature data.
% 这个190是算上HST和TESS数据的总数
When using only HST and TESS data, we obtain an orbital period change rate of $-32.84 \pm 1.78$~ms/yr.
Our findings suggest that a careful selection of observation data can lead to reliable assessments of TTVs of exoplanetary systems. 

%Our findings suggest that a conservative treatment of data and their error bars can lead to reliable assessments of TTVs of exoplanetary systems. 
%and a better understanding of the underlying dynamics .
%Our findings suggest that a conservative treatment of error bars can lead to more reliable assessments of TTV and a better understanding of the underlying dynamics of exoplanetary systems.

To address this issue, we recommend that future transit exoplanet studies should adopt a more cautious approach when reporting error bars for the timing data, and publish the Monte Carlo chain data generated during the light curve fitting together with the light curve data. 
Specifically, researchers should consider the potential impact of incomplete transit coverage and other observational limitations, ensuring that the reported uncertainties reflect a realistic assessment of the observational uncertainties. 
By using more strict standards on error bar reporting, we can enhance the robustness of TTV studies and facilitate comparisons of results from different observational datasets. 
This practice will ultimately contribute to a deeper understanding of the dynamical properties of short period exoplanet systems.

%Our findings suggest that a conservative treatment of error bars can lead to more reliable assessments of TTV and a better understanding of the underlying dynamics of exoplanetary systems. By using more strict standards on error bar reporting, we can enhance the robustness of TTV studies and facilitate comparisons of results from different observational datasets. 
%This practice will ultimately contribute to a deeper understanding of the dynamical properties of short period exoplanet systems.

\section{Conclusion}
In this study, we analyzed transit observations of 37 exoplanets using HST/WFC3 data. By incorporating these high-precision transit times into our linear and quadratic ephemeris models, we identified candidates exhibiting both long- and short-term orbital period variations. Specifically, we classified six planetary systems as long-term orbital period variation candidates, four as short-term variation candidates, and eight as `interesting' candidates that warrant further observational study. 
For many known exoplanets, the uncertainty of their predicted transit windows prohibits an accurate scheduling of follow-up observations. Thus
we also refined the orbital ephemerides of these planets, which will aid in the planning of future observations. Our results demonstrate the importance of high-precision timing data in improving measurements of planetary orbital period change rates. 

% In this study, we have analyzed transit observations of a total of 37 exoplanets from HST/WFC3. 
% By adding these HST/WFC3 transit time data to our linear and quadratic ephemeris model, we are able to identify candidates showing long- and short-term period variation.
% We classify 6 planetary systems as long-term orbital period variation candidates, 4 systems as short-term variation candidates, and 8 as `interesting' candidates, which merit further observation studies. 
% We have also updated the orbital ephemeris of these planets which can be used to construct a better observation plan for these transiting planets in the future. 
% Our results suggest that adding reliable high-precision timing data is crucial to obtain better measurements of the period change rate of planets.
% %By adding these middle transit time data to literature values, we are able to identify candidates showing long- and short-term period variation.
% %We have analyzed transit time data from HST, TESS, and literature for a total of 37 planetary systems, to identify candidates with positive or negative long-term period change rates. 
% %We fit the transit time data using both a linear and a quadratic ephemeris model. 

Our analysis finds strong evidence for orbital period decay in two hot Jupiters. 
Among the roughly 400 known HJs, several appear to have relatively short inspiral timescales. 
Given the older ages of their host stars,  this may suggest a continuous depletion of HJs and thus the need for HJ replenishment to sustain the observed population. These results support the need of a `late-arrived' hot Jupiter population, as proposed by \citet{Chen2023}, which is likely coming through the secular dynamical chaos process \citep{Wu11, Hamers17, Teyssandier19}. 

In the future, we plan to extend this study by incorporating additional space-based transit observation data from JWST \citep{Gardner23}, ET2.0 \citep{Ge2024}, PLATO \citep{Rauer14}, and others. This will allow for more precise estimates of the orbital decay parameter, benefiting from a longer time baseline. The current study can also be applied to the field of short-period transiting brown dwarfs. By investigating the evolution of close brown dwarf companions around solar-type stars, we aim to gain a deeper understanding of the origin of the `brown dwarf desert' \citep{Marcy00, Halbwachs00, Grether06, Ma14}. Furthermore, the HST data presented in this work, when combined with radial velocity data and TTV analysis, can be used to place constraints on potential hidden planets within the system—an important topic in the exoplanet photodynamics research field.

\acknowledgments
We thank our anonymous referee for his/her insightful comments to help improve this manuscript.
We thank \citet{Ivshina22} for compiling the TTV database that served as a valuable resource for our study. We acknowledge the contribution of the NASA Exoplanet Archive, which provided access to a wealth of observational data and resources. 
% 加pylightcurve数据库参数，GAPS Programme (Bonomo et al., 2017) 年龄参数
Furthermore, we extend our appreciation to the HST and TESS mission for its remarkable contribution to exoplanet science. 
%boma: 等审稿意见出来后再看情况加这句。
%We also express our sincere thanks to the anonymous reviewers whose constructive comments and suggestions greatly enhanced the quality of this paper. 
This work has been supported by the National SKA Program of China (grant No. 2022SKA0120101).
BM and CY acknowledge the financial support from the NSFC grant 12073092, the Earth 2.0 mission from SHAO, and the science research grants from the China Manned Space Project  (No. CMS-CSST-2021-B09). LZ acknowledges the financial support of the Postdoctoral Fellowship Program of CPSF (GZB20230767).
JWX and DCC acknowledge the financial support from the NSFC grant (12273011, 12403071), and the National Youth Talent Support Program.

%We acknowledge the financial support from the National Key R\&D Program of China  (2020YFC2201400), NSFC grant 12073092, 12103097, 12103098, the science research grants from the China Manned Space Project  (No. CMS-CSST-2021-B09), and Guangzhou Basic and Applied Basic Research Program   (202102080371).

\vspace{5mm}
\facilities{HST, TESS}

\software{PdotQuest \citep{Wang24}, Iraclis \citep{Tsiaras2016, Tsiaras2016_55, Tsiaras2018}, PyTransit \citep{Parviainen2015}, Emcee \citep{Foreman-Mackey2013}, batman \citep{Kreidberg2015}, ExoTETHyS \citep{2020AJ....159...75M}, Matplotlib \citep{2007CSE.....9...90H}, 
Numpy \citep{oliphant2006guide}, Scipy \citep{2020NatMe..17..261V}, Astropy \citep{AstropyCollaboration2013,AstropyCollaboration2018,AstropyCollaboration2022}
}
%boma: 把汪师姐论文加到引用里， 用 Wang24 代表
%xinyue:

%\bibliography{sample62}{}
\bibliographystyle{aasjournal}
\bibliography{references}

\appendix
\section{Appendix A}
In this section, we show the transit timing data fitting results for all targets showing a constant orbital period trend in our sample, in Figure~\ref{fig:all_plots_appendix} and ~\ref{fig:all_plots_appendix-2}.

\renewcommand{\thefigure}{A\arabic{figure}} % Change figure numbering to A1, A2, etc.
\setcounter{figure}{0}

\begin{figure}[ht!]
\centering
\includegraphics[width=0.85\textwidth]{ 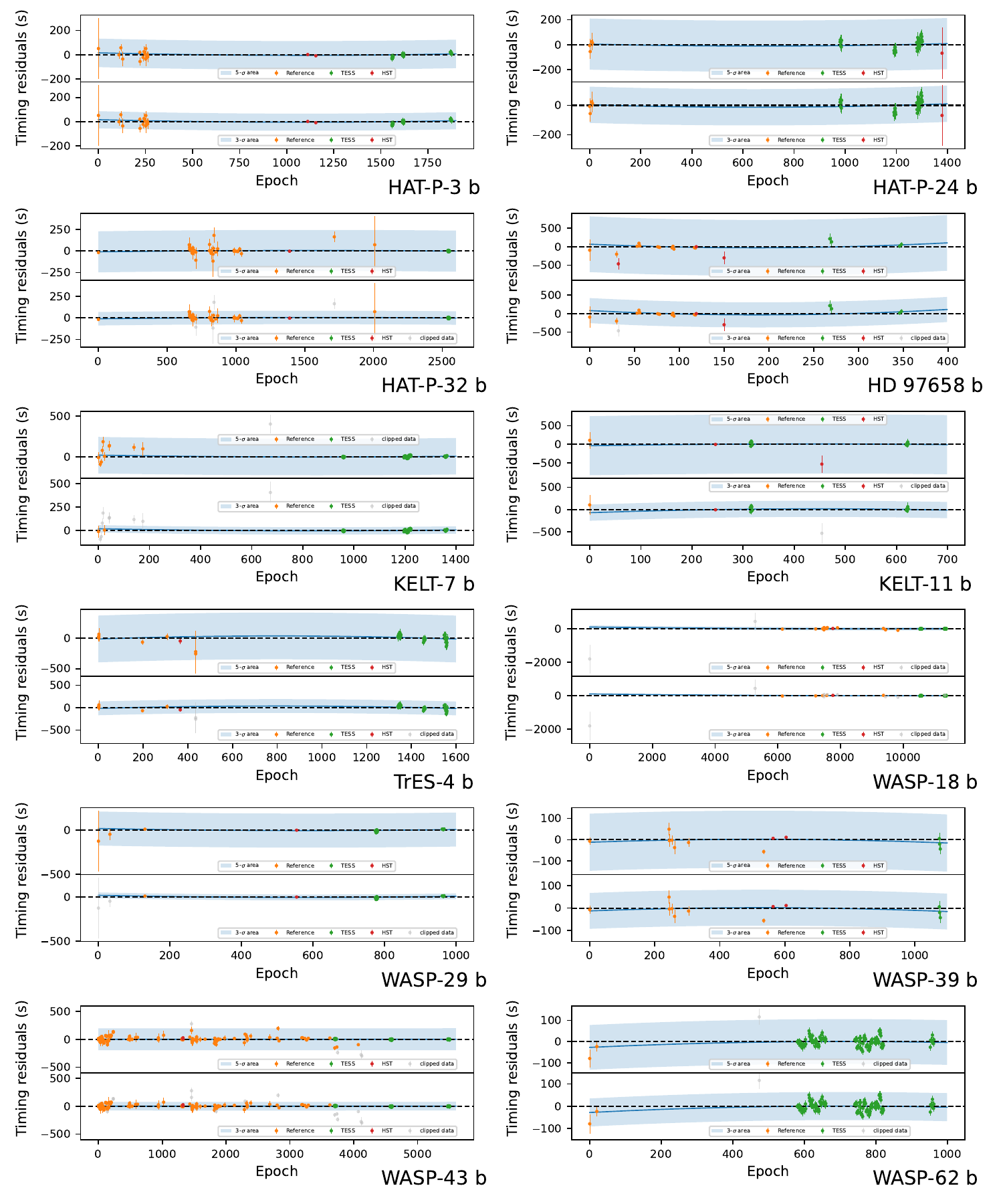}
\caption{
Quadratic TTV model fitting results of all exoplanets in our sample that show an orbital period variation trend consistent with a constant period.
\label{fig:all_plots_appendix}
}
\end{figure}

\begin{figure}[ht!]
\centering
\includegraphics[width=0.85\textwidth]{ 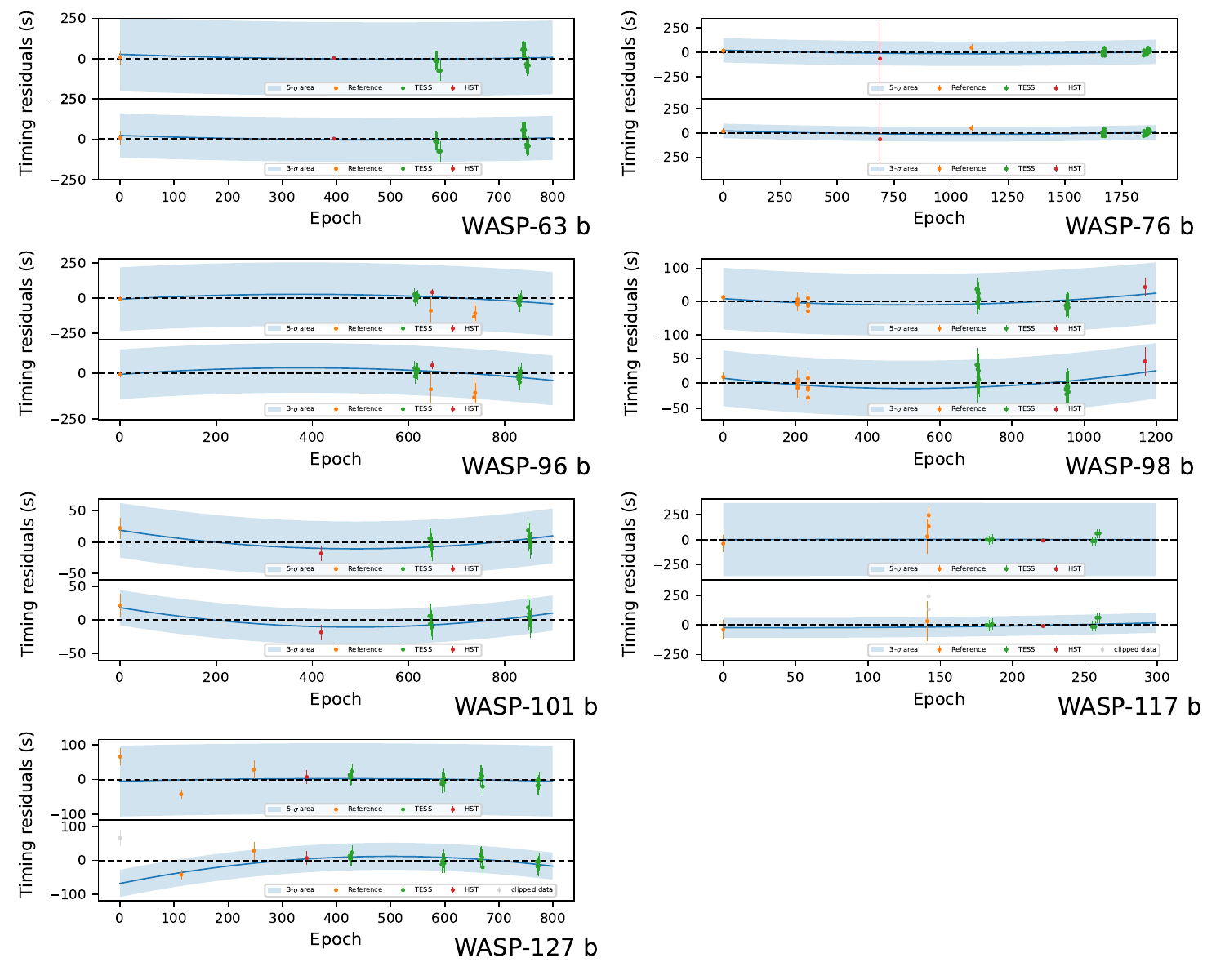}
\caption{
(continued) Quadratic TTV model fitting results of all exoplanets in our sample that show an orbital period variation trend consistent with a constant period.
\label{fig:all_plots_appendix-2}
}
\end{figure}

\section{Appendix B}

Previous literature references for the data used by the 37 exoplanets in this article:

\textbf{WASP-12 b}
\citet{Hebb2009,Chan2011,Sada2012,Cowan2012,Copper2013,Stevenson2014,Kreidberg2015,Maciejewski2016,Collins2017,Patra2017,Patra2020,Wang24}

\textbf{WASP-4 b}
\citet{Wilson2008,Gillon_improved_2009,Winn2009,Dragomir2011,Sanchis2011,Nikolov2012,Hoyer2013,Ranjan2014,Huitson2017,Southworth2019,Baluev2019,Bouma2020,Wang24}

\textbf{WASP-80 b}
\citet{Triaud2013,Mancini2014,Fukui2014,Sedaghati2017,Turner2017,Wang21,Wang24}

\textbf{WASP-19 b}
\citet{Hebb2010,Hellier2011,Dragomir2011,Lendl2013,Tregloan2013,Mancini2013,Espinoza2019,Petrucci20,Wang24} 

\textbf{WASP-17 b}
\citet{Anderson2010,Anderson2011,Southworth2012,Bento2014,Sedaghati_Potassium_2016,Wang24}

\textbf{HAT-P-38 b}
\citet{Mallonn19,Sato2012,Bruno2018,Wang24}

\textbf{HAT-P-2 b}
\citet{Pal2010,Lewis2013,Wang24}

\textbf{HAT-P-11 b}
\citet{Bakos2010,Murgas2019,Wang24}

\textbf{HAT-P-18 b}
\citet{Seeliger2015,Wang24}

\textbf{WASP-178 b}
\citet{Rodriguez2020,Hellier2019,Wang24}

\textbf{WASP-6 b}
\citet{gillon_discovery_2009,Dragomir2011,Sada2012,Jordan2013,Nikolov2015,Tregloan2015,Wang24}

\textbf{HAT-P-26 b}
\citet{Hartman2011_26,Stevenson2016,Wakeford2017,vonEssen2019,Wang24}

\textbf{HAT-P-41 b}
\citet{Hartman2012,Wakeford2020,Wang24}

\textbf{WASP-69 b}
\citet{Bacsturk2019,Wang24}

\textbf{HAT-P-17 b}
\citet{Howard2012,Wang24}

\textbf{WASP-79 b}
\citet{Smalley_WASP7879_2012,Brown2017,Wang24}

\textbf{WASP-121 b}
\citet{Delrez2016,Evans2018,Wang24}

\textbf{HAT-P-12 b}
\citet{Hartman2009,Lee2012,Sada2012,Mallonn2015,Turner2017,Mancini2018,Alexoudi2018,Sariya2021,Wang24}

\textbf{HAT-P-3 b}
\citet{Torres2007,Gibson2010,Chan2011,Wang24}

\textbf{HAT-P-24 b}
\citet{Kipping2010,Wang24}

\textbf{HAT-P-32 b}
\citet{Hartman2011_32_33,2013MNRAS.436.2974G,2014MNRAS.441..304S,2020AAS...23533707Z,Wang24}

\textbf{HD~97658 b}
\citet{Dragomir2013,2020AJ....159..239G,Wang24}

\textbf{KELT-7 b}
\citet{Bieryla2015,Wang24}

\textbf{KELT-11 b}
\citet{Pepper2017,Wang24}

\textbf{TrES-4 b}
\citet{Mandushev2007,2009ApJ...691.1145S,Chan2011,2016MNRAS.459..789T,Wang24}

\textbf{WASP-18 b}
\citet{Hellier2009,2013MNRAS.428.2645M,Wilkins2017,2018MNRAS.477L..21M,Patra2020,Wang24}

\textbf{WASP-29 b}
\citet{Hellier2010,Dragomir2011,Gibson2013,Bonomo2017,Wang24}

\textbf{WASP-39 b}
\citet{Faedi_wasp39_2011,2016ApJ...827...19F,2019AJ....158..144K,Wang24}

\textbf{WASP-43 b}
\citet{Gillon2012,Murgas2014,Chen2014,Stevenson2014,2015PASP..127..143R,Jiang2016,Hoyer2016,Stevenson2017,Patra2020,Wang21,2021MNRAS.508.5514G,2021AJ....162...18S,Wang24}

\textbf{WASP-62 b}
\citet{Hellier2012,Brown2017,Skaf2020,Garhart2020,Wang24}

\textbf{WASP-63 b}
\citet{Hellier2012,Wang24}

\textbf{WASP-76 b}
\citet{West_Three_2016,Ehrenreich2020,Wang24}

\textbf{WASP-96 b}
\citet{Hellier2014,Yip2021,Wang24}

\textbf{WASP-98 b}
\citet{Hellier2014,Mancini2016,Wang24}

\textbf{WASP-101 b}
\citet{Hellier2014,Wang24}

\textbf{WASP-117 b}
\citet{Lendl2014,Mallonn19,Anisman2020,Wang24}

\textbf{WASP-127 b}
\citet{Lam2017,Palle2017,Seidel2020,Skaf2020,Wang24}

\end{document}